\documentclass[conference,final,twoside]{IEEEtran}
\usepackage{cite}

\ifCLASSINFOpdf
   \usepackage[pdftex]{graphicx}
   \graphicspath{{./figures/}{./pdf/}{./jpeg/}}
   \DeclareGraphicsExtensions{.pdf,.jpeg,.png}
\else
   \usepackage[dvips]{graphicx}
   \graphicspath{{./eps/}}
   \DeclareGraphicsExtensions{.eps}
\fi
\usepackage{algorithm}
\usepackage{algpseudocode}

\usepackage[cmex10]{amsmath}
\usepackage{amssymb}

\usepackage[caption=false,font=footnotesize]{subfig}
\usepackage{stfloats}

\usepackage{multirow}

\usepackage{footnote}

\usepackage{threeparttable}

\renewcommand{\vec}[1]{\mathbf{#1}}

\newcommand{\mtx}[1]{\mathbf{#1}}

\begin{document}
%
\title{System-Aware Compression}
%
%
%

\author{\IEEEauthorblockN{Yehuda Dar, Michael Elad, and Alfred M. Bruckstein}
	\IEEEauthorblockA{Computer Science Department, Technion -- Israel Institute of Technology\\
	\{ydar,elad,freddy\}@cs.technion.ac.il}}

%
%

\markboth{}%
{~}
%



\maketitle

\begin{abstract}
Many information systems employ lossy compression as a crucial intermediate stage among other processing components. While the important distortion is defined by the system's input and output signals, the compression usually ignores the system structure, therefore, leading to an overall sub-optimal rate-distortion performance. 
In this paper we propose a compression methodology for an operational rate-distortion optimization considering a known system layout, modeled using linear operators and noise.
Using the alternating direction method of multipliers (ADMM) technique, we show that the design of the new globally-optimized compression reduces to a standard compression of a "system adjusted" signal. 
Essentially, the proposed framework leverages standard compression techniques to address practical settings of the noisy source coding problem. 
We further explain the main ideas of our method by theoretically studying the case of a cyclo-stationary Gaussian signal.
We present experimental results for coding of one-dimensional signals and for video compression using the HEVC standard, showing significant gains by the adjustment to an acquisition-rendering system.
\end{abstract}



%
\IEEEpeerreviewmaketitle

\section{Introduction}
\label{sec:Introduction}

Lossy compression has a central role in information systems where the data may be inaccurately represented in order to meet storage-space or transmission-bandwidth constraints.
While the compression is a crucial system-component, it is only an intermediate stage among data processing procedures that determine the eventual output of the system. 
For example, consider a common audio/visual system structure where the source signal is acquired and compressed for its storage/transmission, then the decompression is followed by a rendering stage producing the ultimate system output. Evidently, in this example, the quality of the system output is determined by the acquisition-rendering chain, and not solely on the lossy compression stage. Nevertheless, the compression is usually designed independently of the system structure, thus inducing a sub-optimal rate-distortion performance for the complete system.

\begin{figure}[]
	\centering
	\includegraphics[width=0.48\textwidth]{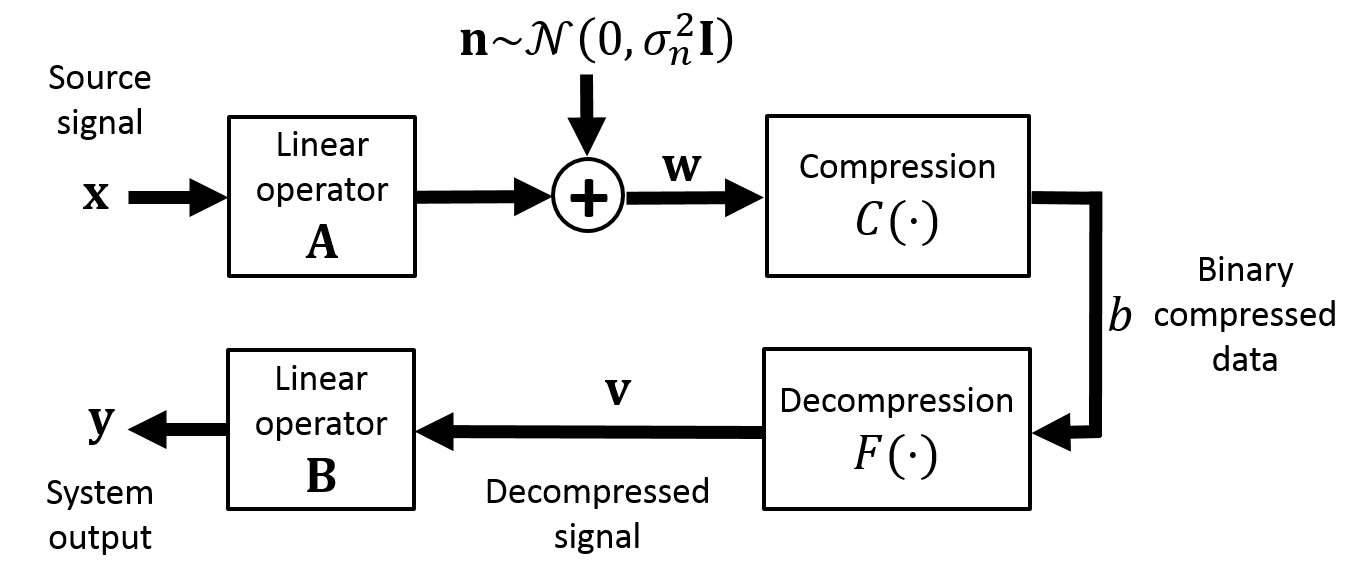}
	\caption{The general system structure considered in this paper.} 
	\label{Fig:general_system_diagram}
\end{figure}

Here we propose a compression approach defined by an operational rate-distortion optimization considering a known system structure. 
Specifically, we study a general flow (Fig. \ref{Fig:general_system_diagram}) where the compression is preceded by a linear operator distorting the input along with an additive white noise, and the decompression is followed by another linear operation.
We formulate a rate-distortion optimization based on a quadratic distortion metric involving the system's linear operators. 
For general linear operators, this intricate rate-distortion optimization is too hard to be directly solved for high dimensional signals.
Consequently, we address this challenge using the alternating direction method of multipliers (ADMM) technique \cite{boyd2011distributed}, suggesting an iterative procedure that relies on the simpler tasks of standard compression (which is system independent!) and $ \ell_2 $-constrained deconvolution for the linear operators of the system.


Optimizing the system output quality from the compression standpoint is an attractive answer to the inherent tradeoff among distortion, bit-cost, and computational complexity. 
First, bits are wisely spent for representing signal components that will be important at the output of the overall system (for example, one should obviously not code signal components belonging to the null space of the post-decompression operator). Second, the added computational load can be well accommodated in the compression environment that is often rich in computational and time resources, in contrast to the decompression stage.

Importantly, the proposed compression framework is a paradigm for addressing intricate rate-distortion optimization forms via iterative solution of easier problems emerging from the ADMM method (or other variable-splitting optimization techniques \cite{dar2017restoration}).
Indeed, the problem addressed here is an extension of our recent compression method \cite{dar2017optimized} that pre-compensates for a later degradation occurring after decompression. In this paper the task is harder than in \cite{dar2017optimized}, as the source signal is available  only in its degraded version. 
The recent wide use of ADMM for complicated signal restoration problems (e.g.,  \cite{afonso2010fast,venkatakrishnan2013plug,dar2016postprocessing,rond2016poisson,romano2017little}) suggests that our ADMM-based approach for complicated compression problems entails great potential.

Essentially, we study here a remote source coding problem (e.g., see its origins in \cite{dobrushin1962information,wolf1970transmission} and their successors). Our contribution is mainly the deterministic setting versus the statistical perspective used in previous works. Specifically, we consider an operational rate-distortion problem for the compression of a given signal, based on a different distortion metric imposed by the lack of an explicit statistical model of the unknown source. 
Our settings lead to the remarkable result that, using the ADMM technique, one can employ a standard compression method to address remote source coding problems in much more complicated instances than were previously feasible (e.g., \cite{al1998lossy}).

Using rate-distortion theory, we further study the examined problem in statistical settings (considering the proposed distortion metric) for the case of a cyclo-stationary Gaussian source signal and linear shift-invariant system operators. Our results show that the initial rate-distortion optimization reduces to a reverse water-filling procedure  adjusted to the system operators and considering the pseudoinverse-filtered version of the input signal. We use these theoretic results to explain concepts appearing (differently) in the proposed practical method intended for non-Gaussian signals and general linear system operators.

Jointly using sampling and source coding procedures is fundamental to digitization-based systems (see, e.g., \cite{kipnis2016distortion}). 
Accordingly, we demonstrate our general framework for adapting standard compression methods to the specific settings of a complete acquisition-rendering system.
We present experiments considering coding of one-dimensional signals using an adaptive tree-based technique, and to video compression using the state-of-the-art HEVC standard \cite{sullivan2012overview}.
Comparisons of our strategy to a regular compression flow exhibited that our method achieves significant gains at medium/high bit-rates.

\section{The Proposed Method}
\label{sec:Proposed Method}

Let us describe the considered system structure (Fig. \ref{Fig:general_system_diagram}). A source signal, an $ N $-length column vector $ \vec{x} \in \mathbb{R}^N $, undergoes a linear processing represented by the $ M \times N $ matrix $ \mtx{A} $ and, then, deteriorated by an additive white Gaussian noise vector $ \vec{n} \sim \mathcal{N} \left( 0, \sigma_n^2 \mtx{I} \right) $, resulting in the signal 
\begin{IEEEeqnarray}{rCl}
	\label{eq:compression input - w - definition}
	\vec{w} = \mtx{A} \vec{x} + \vec{n} 
\end{IEEEeqnarray}
where $ \vec{w} $ and $ \vec{n} $ are $ M $-length column vectors. 
We represent the lossy compression procedure via the mapping $ C: \mathbb{R}^M \rightarrow \mathcal{B} $ from the $ M $-dimensional signal domain to a discrete set $ \mathcal{B} $ of binary compressed representations (that may have different lengths).
The signal $ \vec{w} $ is the input to the compression component of the system, producing the compressed binary data $\textit{b} = C \left( \vec{w} \right)$ that can be stored or transmitted in an error-free manner.
Then, on a device and settings depending on the specific application, the compressed data $ \textit{b} \in \mathcal{B} $ is decompressed to provide the signal $\vec{v} = F \left( \textit{b} \right)$ where $ F: \mathcal{B} \rightarrow \mathcal{S} $ represents the decompression mapping between the binary compressed representations in $ \mathcal{B} $ to the corresponding decompressed signals in the discrete set $ \mathcal{S} \subset \mathbb{R}^M $.
The decompressed signal $ \vec{v} $ is further processed by the linear operator denoted as the $ N \times M $ matrix $ \mtx{B} $, resulting in the system output signal 
\begin{IEEEeqnarray}{rCl}
	\label{eq:system output - y - definition}
	\vec{y} = \mtx{B} \vec{v}, 
\end{IEEEeqnarray}
which is an $ N $-length real-valued column vector.

As an example, consider an acquisition-compression-rendering system where the signal $ \vec{w} $ is a sampled version of the source signal $ \vec{x} $, and the system output $ \vec{y} $ is the rendered version of the decompressed signal $ \vec{v} $.

We assume here that the operators $ \mtx{A} $ and $ \mtx{B} $, as well as the noise variance $ \sigma_n^2 $, are known and fixed (i.e., cannot be optimized). Consequently, we formulate a new compression procedure in order to optimize the end-to-end rate-distortion performance of the entire system.
Specifically, we want the system output $ \vec{y} $ to be the best approximation of the source signal $ \vec{x} $ under the bit-budget constraint. However, at the compression stage we do not accurately know $ \vec{x} $, but rather its degraded form $ \vec{w} $ formulated in (\ref{eq:compression input - w - definition}). This motivates us to suggest the following distortion metric with respect to the system output $ \vec{y} $
\begin{IEEEeqnarray}{rCl}
	\label{eq:suggested distortion - system}
	d_{s}\left( \vec{w}, \vec{y} \right) =  \frac{1}{M} \left\| { \vec{w} - \mtx{A} \vec{y} } \right\|_2^2 .
\end{IEEEeqnarray}
This metric conforms with the fact that if $ \vec{y} $ is close to $ \vec{x} $, 
then, by (\ref{eq:compression input - w - definition}), $\vec{w}$ will be close to $\mtx{A} \vec{y}$ up to the noise $ \vec{n} $.
Indeed, for the ideal case of $ \vec{y} = \vec{x} $ the metric (\ref{eq:suggested distortion - system}) becomes 
\begin{IEEEeqnarray}{rCl}
	\label{eq:suggested distortion  - system - ideal solution}
	d_{s}\left( \vec{w}, \vec{x} \right) =  \frac{1}{M} \left\| { \vec{n} } \right\|_2^2 \approx \sigma_n^2
\end{IEEEeqnarray}
where the last approximate equality is under the assumption of a sufficiently large $ M $ (the length of $ \vec{n} $). 
Since $ \vec{y} = \mtx{B} \vec{v} $, we can rewrite the distortion $ d_{s}\left( \vec{w}, \vec{y} \right) $ in (\ref{eq:suggested distortion - system}) as a function of the decompressed signal $ \vec{v} $, namely, 
\begin{IEEEeqnarray}{rCl}
	\label{eq:suggested distortion - compression}
	d_{c}\left( \vec{w}, \vec{v} \right) =  \frac{1}{M} \left\| { \vec{w} - \mtx{A} \mtx{B}\vec{v} } \right\|_2^2 .
\end{IEEEeqnarray}
Since the operator $ \mtx{B} $ produces the output signal $ \vec{y} $, an ideal result will be $ \vec{y} = \mtx{P}_B \vec{x} $, where $ \mtx{P}_B $ is the matrix projecting onto $ \mtx{B} $'s range. The corresponding ideal distortion is 
\begin{IEEEeqnarray}{rCl}
	\label{eq:suggested distortion  - compression - ideal solution}
	D_0 \triangleq d_{s}\left( \vec{w},  \mtx{P}_B\vec{x} \right)  = \frac{1}{M} \left\| { \mtx{A} \left( \mtx{I} - \mtx{P}_B \right) \vec{x} + \vec{n} } \right\|_2^2 .
\end{IEEEeqnarray}

We use the distortion metric (\ref{eq:suggested distortion - compression}) to constrain the bit-cost minimization in the following rate-distortion optimization
\begin{IEEEeqnarray}{rCl}
	\label{eq:rate-distortion optimization - constrained}
	\begin{aligned}
		& \hat{ \vec{v}} = \underset{ \vec{v}\in\mathcal{S} }{\text{argmin}}
		& & { R \left( \vec{v} \right) } \\
		& \text{subject to}
		& & D_{0} \le \frac{1}{M} \left\| { \vec{w} - \mtx{A} \mtx{B}\vec{v} } \right\|_2^2 \le D_{0} + D 
	\end{aligned}
\end{IEEEeqnarray}
where $R \left( \vec{v} \right)$ evaluates the length of the binary compressed description of the decompressed signal $ \vec{v} $, and $ D \ge 0 $ determines the allowed distortion. 
By (\ref{eq:suggested distortion  - compression - ideal solution}), the value $ D_{0} $ depends on the operator $ \mtx{A} $, the null space of $ \mtx{B} $, the source signal $ \vec{x} $, and the noise realization $ \vec{n} $. 
Since $ \vec{x} $ and $ \vec{n} $ are unknown, $ D_{0} $ cannot be accurately calculated in the operational case (in Section \ref{sec:Theory} we formulate the expected value of $ D_{0} $ for the case of a cyclo-stationary Gaussian source signal).
We address the optimization (\ref{eq:rate-distortion optimization - constrained}) using its unconstrained Lagrangian form 
\begin{IEEEeqnarray}{rCl}
	\label{eq:rate-distortion optimization - Lagrangian}
	\hat{ \vec{v}} = \underset{ \vec{v}\in\mathcal{S} }{\text{argmin}}
	~~ { R \left( \vec{v} \right) + \lambda \frac{1}{M} \left\| { \vec{w} - \mtx{A} \mtx{B}\vec{v} } \right\|_2^2 }
\end{IEEEeqnarray}
where $ \lambda \ge 0 $ is a Lagrange multiplier corresponding to some distortion constraint $ D_{\lambda} \ge D_{0} $ (such optimization strategy with respect to some Lagrange multiplier is common, e.g., in video coding \cite{sullivan2012overview}).
In the case of high-dimensional signals, the discrete set $ \mathcal{S} $ is extremely large and, therefore, it is impractical to directly solve the Lagrangian form in (\ref{eq:rate-distortion optimization - Lagrangian}) for generally structured matrices $ \mtx{A} $ and $ \mtx{B} $. 
This difficulty vanishes, for example, when $ \mtx{A} = \mtx{B} = \mtx{I} $, reducing the Lagrangian optimization in (\ref{eq:rate-distortion optimization - Lagrangian}) to the standard (system independent) compression form (see, e.g., \cite{shoham1988efficient,ortega1998rate}). Indeed, such standard Lagrangian rate-distortion optimizations are practically solved using block-based designs that translate the task to a sequence of block-level optimizations of feasible dimensions.

Here we consider general $ \mtx{A} $ and $ \mtx{B} $ matrices, and address the computational difficulty in solving (\ref{eq:rate-distortion optimization - Lagrangian}) using the alternating direction method of multipliers (ADMM) technique \cite{boyd2011distributed}.
For start, we apply variable splitting to rewrite (\ref{eq:rate-distortion optimization - Lagrangian}) as 
\begin{IEEEeqnarray}{rCl}
	\label{eq:rate-distortion optimization - variable splitting}
	\begin{aligned}
		& \hat{ \vec{v}} = \underset{ \vec{v}\in\mathcal{S} , {\vec{z}}\in\mathbb{R}^M }{\text{argmin}}
		~~ { R \left( \vec{v} \right) + \lambda \frac{1}{M} \left\| { \vec{w} - \mtx{A} \mtx{B}\vec{z} } \right\|_2^2 } \\
		& \text{subject to} ~~~~ \vec{v} = \vec{z}
	\end{aligned}
\end{IEEEeqnarray}
where $ \vec{z} \in \mathbb{R}^M $ is an auxiliary variable that is not (directly) restricted to the discrete set $ \mathcal{S} $.
The augmented Lagrangian (in its scaled form) and the method of multipliers (see \cite[Ch. 2]{boyd2011distributed}) turn (\ref{eq:rate-distortion optimization - variable splitting}) into the following iterative procedure 
\begin{IEEEeqnarray}{rCl}
	\label{eq:rate-distortion optimization - augmented Lagrangian}
	&& \left( \hat{ \vec{v}}^{(t)}, \hat{\vec{z}}^{(t)} \right) = 
	\\ \nonumber
	&& \mathop {{\text{argmin}}}\limits_{ \vec{v}\in\mathcal{S} , {\vec{z}}\in\mathbb{R}^M }  R \left( \vec{v} \right) + \lambda \frac{1}{M} \left\| { \vec{w} - \mtx{A} \mtx{B}\vec{z} } \right\|_2^2 + \frac{\beta}{2}{\left\| { \vec{v} - \vec{z} + \vec{u}^{(t)} } \right\|_2^2} 
	\\ 
	&& \vec{u}^{(t+1)} = \vec{u}^{(t)} + \left( \hat{ \vec{v}}^{(t)} - \hat{\vec{z}}^{(t)} \right),
\end{IEEEeqnarray}
where $ t $ is the iteration number, $\vec{u}^{(t)} \in \mathbb{R}^M$ is the scaled dual variable, and $ \beta $ is an auxiliary parameter originating at the augmented Lagrangian.
Further simplifying (\ref{eq:rate-distortion optimization - augmented Lagrangian}) using one iteration of alternating minimization gives the ADMM form of the problem 
\begin{IEEEeqnarray}{rCl}
	\label{eq:rate-distortion optimization - ADMM - compression}
	&& \hat{\vec{v}}^{(t)} = \mathop {{\text{argmin}}}\limits_{ \vec{v}\in\mathcal{S} }  R \left( \vec{v} \right) + \frac{\beta}{2}{\left\| {  \vec{v} - \tilde{ \vec{z}}^{(t)} } \right\|_2^2}
	\\
	\label{eq:rate-distortion optimization - ADMM - deconvolution}
	&& \hat{ \vec{z}}^{(t)} = \mathop {\text{argmin}}\limits_{{\vec{z}}\in\mathbb{R}^M } \lambda \frac{1}{M} \left\| { \vec{w} - \mtx{A} \mtx{B} \vec{z} } \right\|_2^2 + \frac{\beta}{2}{\left\| {  \vec{z} - \tilde{ \vec{v}}^{(t)} } \right\|_2^2}
	\\
	\label{eq:rate-distortion optimization - ADMM - u update}
	&& \vec{u}^{(t+1)} = \vec{u}^{(t)} + \left( \hat{ \vec{v}}^{(t)} - \hat{\vec{z}}^{(t)} \right).
\end{IEEEeqnarray}
where $ \tilde{ \vec{z}}^{(t)} = \hat{\vec{z}}^{(t-1)} - \vec{u}^{(t)} $ and $ \tilde{ \vec{v}}^{(t)} = \hat{\vec{v}}^{(t)} + \vec{u}^{(t)} $. 
Importantly, the compression design, expressed by $ \left\lbrace \mathcal{S}, R \right\rbrace $, and the system-specific operators  $ \left\lbrace \mtx{A}, \mtx{B} \right\rbrace $ were decoupled by the ADMM to reside in distinct optimization problems that are easier to solve.

We identify the first stage (\ref{eq:rate-distortion optimization - ADMM - compression}) as the Lagrangian optimization employed for standard compression (and decompression) tasks considering the regular mean squared error metric. Specifically, the Lagrange multiplier for this standard optimization is $ \tilde{\lambda} = \frac{\beta M}{2 } $.
Furthermore, we suggest to replace the solution of (\ref{eq:rate-distortion optimization - ADMM - compression}) with the application of a standard compression (and decompression) technique, even one that does not follow the Lagrangian optimization formulated in (\ref{eq:rate-distortion optimization - ADMM - compression}).
We denote the standard compression and decompression via 
\begin{IEEEeqnarray}{rCl}
	\label{eq:rate-distortion optimization - ADMM - compression - standard compression}
	{\textit{b}}^{(t)} = StandardCompress \left( \tilde{ \vec{z}}^{(t)}, \theta \right)
	\\
	\label{eq:rate-distortion optimization - ADMM - compression - standard decompression}
	\hat{\vec{v}}^{(t)} = StandardDecompress \left( {\textit{b}}^{(t)} \right)		
\end{IEEEeqnarray}
where $ \theta $ is a general parameter that extends the Lagrange multiplier role in determining the rate-distortion tradeoff (see Algorithm \ref{Algorithm:Proposed Method}). This important suggestion defines our method as a generic approach that can optimize any compression technique with respect to the specific system it resides in.

The second optimization stage (\ref{eq:rate-distortion optimization - ADMM - deconvolution}) is an $ \ell_2 $-constrained deconvolution problem, that can be easily solved in various ways. The analytic solution of (\ref{eq:rate-distortion optimization - ADMM - deconvolution}) is 
\begin{IEEEeqnarray}{rCl}
	\label{eq:rate-distortion optimization - ADMM - deconvolution - analytic solution}
	\hat{ \vec{z}}^{(t)} = \left(  \mtx{B}^{*} \mtx{A}^{*} \mtx{A} \mtx{B} + \frac{\beta M}{2\lambda} \mtx{I}  \right)^{-1} \left( \mtx{B}^{*} \mtx{A}^{*} \vec{w} + \frac{\beta M}{2\lambda} \tilde{ \vec{v}}^{(t)}  \right),~~~  
\end{IEEEeqnarray}
showing it as a weighted averaging of $ \vec{w} $ and $ \tilde{ \vec{v}}^{(t)}  $. In the generic description given in Algorithm \ref{Algorithm:Proposed Method} we replace the quantity $ \frac{\beta M}{2\lambda} $ with the parameter $ \tilde{\beta} $.

The proposed method is summarized in Algorithm \ref{Algorithm:Proposed Method}. 
Our goal is to provide a binary compressed representation of the optimized solution. Hence, the procedure output is the compressed data, ${\textit{b}}^{(t)}$, obtained in the compression stage of the last iteration.

\begin{algorithm}
	\caption{Generic System-Aware Compression}
	\label{Algorithm:Proposed Method}
	\begin{algorithmic}[1]
		\State Inputs: $ \vec{w} $, $ \theta $, $ \tilde{\beta} $.
		\State  Initialize $t = 0$ , $ {\hat{\vec{z}}}^{(0)} = \vec{w} $ , $\vec{u}^{(1)} = \vec{0}$.
		\Repeat 

		\State $ t \gets t + 1$
				
		\State $ \tilde{ \vec{z}}^{(t)} = \hat{\vec{z}}^{(t-1)} - \vec{u}^{(t)} $

		\State $ {\textit{b}}^{(t)} = StandardCompress \left( \tilde{ \vec{z}}^{(t)}, \theta \right) $
		\State $ \hat{\vec{v}}^{(t)} = StandardDecompress \left( {\textit{b}}^{(t)} \right) $
		
		\State $ \tilde{ \vec{v}}^{(t)} = \hat{\vec{v}}^{(t)} + \vec{u}^{(t)} $
		\State $\hat{ \vec{z}}^{(t)} = \left(  \mtx{B}^{*} \mtx{A}^{*} \mtx{A} \mtx{B} + \tilde{\beta} \mtx{I}  \right)^{-1} \left( \mtx{B}^{*} \mtx{A}^{*} \vec{w} + \tilde{\beta} \tilde{ \vec{v}}^{(t)}  \right)$
		
		\State $\vec{u}^{(t+1)} = \vec{u}^{(t)} + \left( \hat{ \vec{v}}^{(t)} - \hat{\vec{z}}^{(t)} \right)$
		
		\Until{stopping criterion is satisfied}
		\State Output: $ {\textit{b}}^{(t)} $, which is the binary compressed data obtained in the last iteration.
	\end{algorithmic}
\end{algorithm}

\section{Theoretic Analysis for the Gaussian Case}
\label{sec:Theory}

In this section we use the rate-distortion theory framework to further explore the fundamental problem of system-optimized compression. We consider the case of a cyclo-stationary Gaussian signal and system involving linear shift-invariant operators and no noise, yet, the obtained results exhibit the prominent ideas of the general problem and the operational method presented in Section \ref{sec:Proposed Method}. 

\subsection{Problem Formulation and Solution}
\label{subsec:Theory - formulation and solution}
The source signal is modeled here as $\vec{x} \sim  \mathcal{N}\left(0,\mtx{R}_{\vec{x}}\right) $, i.e., a zero-mean Gaussian random vector with a circulant autocorrelation matrix $ \mtx{R}_{\vec{x}} $. The eigenvalues of $ \mtx{R}_{\vec{x}} $ are denoted as $\left\lbrace \lambda_k^{\left( \vec{x}\right)} \right\rbrace_{k=0}^{N-1}$.
The first processing part of the system produces the signal ${\vec{w}} = \mtx{A} \vec{x}$, where here $ \mtx{A} $ is a real-valued $N\times N$ circulant matrix. 
Evidently, the signal $ \vec{w} $ is a zero-mean Gaussian random vector with autocorrelation matrix $ \mtx{R}_{\vec{w}}\nolinebreak=\nolinebreak\mtx{A}  \mtx{R}_{\vec{x}} \mtx{A}^* $.

Here $ \mtx{A} $ and $ \mtx{B} $ are circulant $ N \times N $ matrices, thus, diagonalized by the $ N\times N $ Discrete Fourier Transform (DFT) matrix 
$ \mtx{F} $ as $\mtx{F} \mtx{A} \mtx{F}^* = \mtx{\Lambda}_{\mtx{A}} $ and $ \mtx{F} \mtx{B} \mtx{F}^* = \mtx{\Lambda}_{\mtx{B}} $, where $ \mtx{\Lambda}_{\mtx{A}} $ and $ \mtx{\Lambda}_{\mtx{B}} $ are diagonal matrices formed by the elements $ \left\lbrace a^F_k \right\rbrace_{k=0}^{N-1} $ and $ \left\lbrace b^F_k \right\rbrace_{k=0}^{N-1} $, respectively. 
Accordingly, the pseudoinverse matrix of $ \mtx{A} $ is defined as $\mtx{A}^{+} =  \mtx{F}^* \mtx{\Lambda}_{\mtx{A}}^{+} \mtx{F} $, where $ \mtx{\Lambda}_{\mtx{A}}^{+} $ is the pseudoinverse of $ \mtx{\Lambda}_{\mtx{A}} $, an $ N \times N $ diagonal matrix with the $ k^{th} $ diagonal component $ a^{F,+}_k = {1}/{a^F_k} $ for $ k $ where $ a^F_k\ne 0 $, and $ a^{F,+}_k = 0 $ for $ k $ where $ a^F_k = 0 $.
The matrix $ \mtx{B} $ has corresponding definitions to those given above for $ \mtx{A} $.

Recall that the system output is $ \vec{y} = \mtx{B} \vec{v} $, where here $ \vec{v} $ is the random vector representing the decompressed signal. Accordingly, in the theoretic framework here, the rate is associated with the mutual information $ I\left( \vec{w}, \vec{y}\right) $.
In (\ref{eq:rate-distortion optimization - constrained}) we formulated the operational rate-distortion optimization describing our practical task, cast here in the theoretic framework to 
\begin{IEEEeqnarray}{rCl}
	\label{eq:theory - rate-distortion optimization - basic}
	\begin{aligned}
		& \underset{ p_{\vec{v}|\vec{w}} }{\text{min}}
		& & { I \left( \vec{w}; \mtx{B}\vec{v} \right) } \\
		& \text{s.t.}
		& & N  E \left\lbrace D_0 \right\rbrace  \le E\left\lbrace \left\| { \vec{w} - \mtx{A} \mtx{B}\vec{v} } \right\|_2^2 \right\rbrace \le N \left(  E \left\lbrace D_0 \right\rbrace + D \right)
	\end{aligned}
\end{IEEEeqnarray}
where $ p_{\vec{v}|\vec{w}} $ is the conditional probability-density-function of $ \vec{v} $ given $ \vec{w} $, and $ D \ge 0 $ determines the allowed expected distortion. The value $E \left\lbrace D_0 \right\rbrace$, stemming from (\ref{eq:suggested distortion  - compression - ideal solution}) and the noiseless settings here, is the minimal expected distortion evaluated (see Appendix \ref{Appendix:theory:subsec:Minimal Expected Distortion}) as 
\begin{IEEEeqnarray}{rCl}
	\label{eq:theory - rate-distortion optimization - expected D_0}
	E \left\lbrace D_0 \right\rbrace = \frac{1}{N} \sum\limits_{k: a_k^F \ne 0 ~,~ b_k^F = 0} {\left\lvert a_k^F \right\rvert ^2 \lambda_k^{ \left( \vec{x} \right) }} .
\end{IEEEeqnarray}

We state that the basic rate-distortion optimization problem in (\ref{eq:theory - rate-distortion optimization - basic}) is equivalent to (see proof in Appendix \ref{Appendix:theory:subsec:Equivalence of Optimization Problems 1 and 2})
\begin{IEEEeqnarray}{rCl}
	\label{eq:theory - rate-distortion optimization - pseudoinverse filtered input}
	\begin{aligned}
		& \underset{ p_{\vec{v}|\tilde{\vec{w}}} }{\text{min}}
		& & { I \left( \tilde{\vec{w}}; \mtx{P}_{\mtx{B}} \mtx{P}_{\mtx{A}} \vec{v} \right) } \\
		& \text{s.t.}
		& & E\left\lbrace \left\| { \mtx{A} \mtx{B} \left( \tilde{\vec{w}} - \vec{v} \right) } \right\|_2^2 \right\rbrace \le N D  ~~~~~ 
	\end{aligned}
\end{IEEEeqnarray}
where $ {\tilde{\vec{w}}} = \mtx{B}^{+} \mtx{A}^{+} \vec{w} $ is the pseudoinverse filtered compression-input $ \vec{w} $. Here $ \mtx{P}_{\mtx{A}} $ and $ \mtx{P}_{\mtx{B}} $ are projection matrices corresponding to the range of $ \mtx{A} $ and $ \mtx{B} $, respectively.
We define the set $ \mathcal{K}_{AB} \nolinebreak \triangleq \nolinebreak \left\lbrace  k : a_k^F \ne 0 ~\text{and}~ b_k^F \ne 0 \right\rbrace $ that contains the DFT-domain component indices belonging to the range of the joint operator $ \mtx{AB} $.

Since $ \tilde{\vec{w}} $ is a cyclo-stationary Gaussian random vector, we transform the optimization (\ref{eq:theory - rate-distortion optimization - pseudoinverse filtered input}) into a Fourier-domain distortion-allocation problem considering independent Gaussian variables (see proof in Appendix \ref{Appendix:theory:subsec:Equivalence of Optimization Problems 2 and 3}), namely, 
\begin{IEEEeqnarray}{rCl}
	\label{eq:theoretic Gaussian analysis - DFT-domain - Gaussian distortion allocation - general}
	\begin{aligned}
		& \underset{\left\lbrace D_k \right\rbrace_{k \in \mathcal{K}_{AB}}}{\text{min}}
		& & \sum\limits_{k \in \mathcal{K}_{AB}} { \frac{1}{2} \log \left( { \frac{ \lambda^{\left(\tilde{\vec{w}}\right)}_k   }{ D_k } } \right)  } \\
		& \text{s.t.}
		& & \sum\limits_{k \in \mathcal{K}_{AB}} { \left| a^F_k b^F_k\right|^2 D_k } \le N D ~~~
		\\
		& & & 0 \le D_k \le \lambda^{\left(\tilde{\vec{w}}\right)}_k ~~~,~{k \in \mathcal{K}_{AB}}.
	\end{aligned}
\end{IEEEeqnarray}
where for ${k \in \mathcal{K}_{AB}}$ the value $ \lambda^{\left(\tilde{\vec{w}}\right)}_k = \lambda^{\left({\vec{w}}\right)}_k / {\left| a^F_k b^F_k\right|^2} $ is the variance of the $ k^{th} $ DFT-coefficient of $ {\tilde{\vec{w}}} $.
The solution of problem (\ref{eq:theoretic Gaussian analysis - DFT-domain - Gaussian distortion allocation - general}), obtained using Lagrangian optimization and the KKT conditions, is given by the following distortion allocation: for $ k\in \mathcal{K}_{AB} $ 
\begin{IEEEeqnarray}{rCl}
	\label{eq:theoretic Gaussian analysis - DFT-domain - Gaussian distortion allocation - optimal distortions}
	D_k = 
	 \left\{ {\begin{array}{*{20}{c}}
			{\theta / \left| a^F_k b^F_k \right|^2}&{,\text{for~~} 0 \le \theta < \left| a^F_k b^F_k \right|^2 \lambda^{\left(\tilde{\vec{w}}\right)}_k  } \\ 
			\lambda^{\left(\tilde{\vec{w}}\right)}_k  & {,\text{for~~~} \theta \ge \left| a^F_k b^F_k \right|^2 \lambda^{\left(\tilde{\vec{w}}\right)}_k   ,~~ } 
		\end{array}} \right.  \nonumber\\ 
\end{IEEEeqnarray}
where $ \theta $ is set such that $ \sum\limits_{k \in \mathcal{K}_{AB}} { \left| a^F_k b^F_k\right|^2 D_k } = N D $ is satisfied. Importantly, for $ k\notin \mathcal{K}_{AB} $ the rate is $ R_k = 0 $ and the distortion is set to $ D_k = 0 $ (see Appendix \ref{Appendix:theory:subsec:Equivalence of Optimization Problems 2 and 3}). 
The optimal rates corresponding to (\ref{eq:theoretic Gaussian analysis - DFT-domain - Gaussian distortion allocation - optimal distortions}) assign for $ k\in \mathcal{K}_{AB} $ that also obeys $ 0 \le \theta < \left| a^F_k b^F_k \right|^2 \lambda^{\left(\tilde{\vec{w}}\right)}_k  $ \vspace{-0.25cm}
\begin{IEEEeqnarray}{rCl}
	\label{eq:theoretic Gaussian analysis - DFT-domain - Gaussian distortion allocation - optimal rates}
	R_k =\frac{1}{2} \log \left(  { \left| a^F_k b^F_k \right|^2 \frac{ \lambda^{\left(\tilde{\vec{w}}\right)}_k }{\theta}  } \right)
\end{IEEEeqnarray}
and otherwise $ R_k = 0 $. 
Eq. (\ref{eq:theoretic Gaussian analysis - DFT-domain - Gaussian distortion allocation - optimal rates}) shows that the optimal rate assignments include compensation for the modulation applied beforehand in the pseudoinverse filtering of the input $ \vec{w} $. Moreover, DFT components belonging to the null spaces of $ \mtx{A} $ and $ \mtx{B} $ are not coded (i.e., get zero rates).

\begin{figure*}[ht!]
	\centering
	{\subfloat[]{\label{fig:experiments_1d_result_regular}\includegraphics[width=0.31\textwidth]{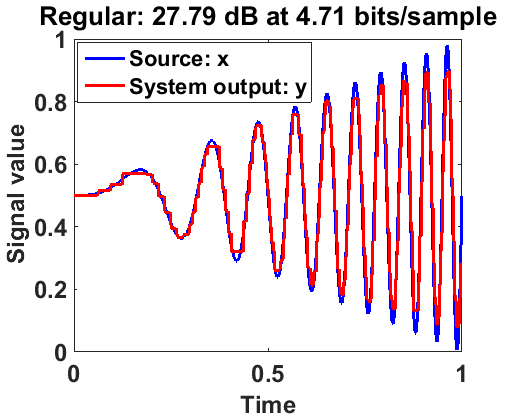}}}
	{\subfloat[]{\label{fig:experiments_1d_result_proposed}\includegraphics[width=0.31\textwidth]{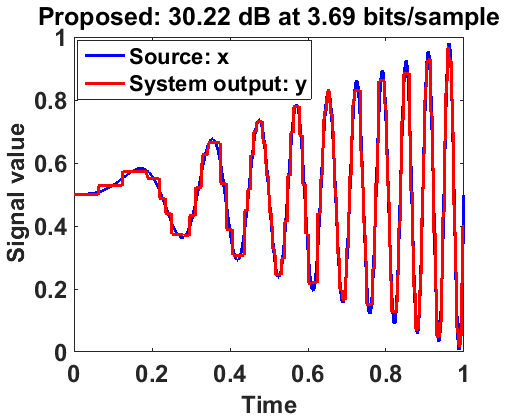}}}
	{\subfloat[]{\label{fig:experiments_1d_psnr_rate_curves}\includegraphics[width=0.31\textwidth]{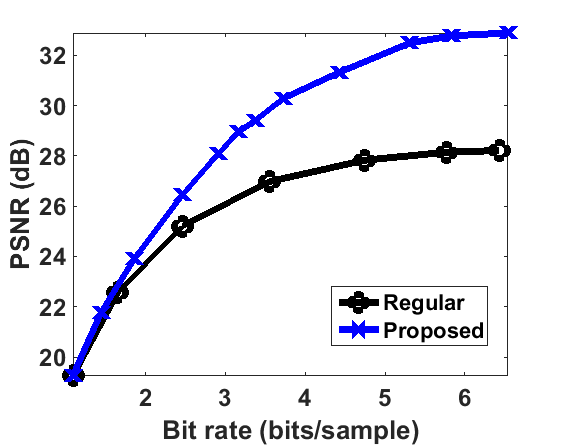}}}
	\caption{Experiment for one-dimensional signal and adaptive tree-based coding in an acquisition-rendering system. The amplitude-modulated chirp source signal is the blue curve in (a) and (b). The red curves in (a) and (b) are the system output (rendered) signals resulting from regular compression and the proposed method, respectively. (c) Comparison of PSNR-bitrate curves.}
	\label{Fig:Experimental Results - 1d signals - chirp}
\end{figure*}

\subsection{Problem Solution in Theory and Practice }
\label{subsec:Theory - Theory and Practice}

While the theoretic framework above considers a cyclo-stationary Gaussian signal and a noiseless system composed of linear shift-invariant operators, the solution presented exhibits important ideas that also appear in the practical method of Section \ref{sec:Proposed Method}. Recall that our method is designed for the operational settings of the problem treating non-Gaussian signals and general linear operators, hence, the resemblances between the above theory and practice are at the conceptual level and may materialize differently.

We addressed the theoretic problem (\ref{eq:theory - rate-distortion optimization - basic}) using a simple inverse filtering of the input data $ \vec{w} $, transforming the problem into (\ref{eq:theory - rate-distortion optimization - pseudoinverse filtered input}) and (\ref{eq:theoretic Gaussian analysis - DFT-domain - Gaussian distortion allocation - general}) that were solved using an extended version of the standard reverse water-filling procedure. 
Analogously, our practical method (Algorithm \ref{Algorithm:Proposed Method}) repeatedly compresses a signal formed by an $ \ell_2 $-constrained deconvolution filtering, that can be rewritten also as a pseudoinverse filtering of the input followed by a weighted averaging with $ \tilde{ \vec{v}}^{(t)} $, i.e., 
\begin{IEEEeqnarray}{rCl}
	\label{eq:theoretic Gaussian analysis - theory and practice - constrained deconvolution interpretation}
	\hat{ \vec{z}}^{(t)} = \left(  \mtx{B}^{*} \mtx{A}^{*} \mtx{A} \mtx{B} + \tilde{\beta} \mtx{I}  \right)^{-1} \left( \mtx{B}^{*} \mtx{A}^{*} \mtx{A} \mtx{B}  \tilde{\vec{w}} + \tilde{\beta} \tilde{ \vec{v}}^{(t)}  \right).~~~
\end{IEEEeqnarray}
This shows that, in practice as well as in theory, the input $ \vec{w} $ should go through a pseudoinverse filtering (softened via (\ref{eq:theoretic Gaussian analysis - theory and practice - constrained deconvolution interpretation})) as a preceding stage to compression. 

The second prominent principle of the theoretic solution, exhibited in (\ref{eq:theoretic Gaussian analysis - DFT-domain - Gaussian distortion allocation - optimal rates}), is to compensate for the modulation applied by the pseudoinverse filter corresponding to the effective system operator $ \mtx{A} \mtx{B} $. 
Similarly in Algorithm \ref{Algorithm:Proposed Method}, the constrained deconvolution stage (\ref{eq:theoretic Gaussian analysis - theory and practice - constrained deconvolution interpretation}) implements this idea by better preserving $ \tilde{\vec{w}} $ components corresponding to higher energy parts of $ \mtx{A} \mtx{B} $.  This is clearly observed in the particular case of circulant $ \mtx{A} $ and $ \mtx{B} $, where the filtering (\ref{eq:theoretic Gaussian analysis - theory and practice - constrained deconvolution interpretation}) reduces to the DFT-domain component-level operation of 
\begin{IEEEeqnarray}{rCl}
	\label{eq:theoretic Gaussian analysis - theory and practice - constrained deconvolution interpretation - DFT for circulant operators}
	\hat{ {z}}^{F,(t)}_k = \left(  \left| a^F_k b^F_k \right|^2 \tilde{{w}}^{F}_k + \tilde{\beta} \tilde{ {v}}^{F,(t)}_k \right) / \left(  { \left| a^F_k b^F_k \right|^2 + \tilde{\beta} }  \right)
\end{IEEEeqnarray}
where $ \tilde{{w}}^{F}_k $ and $ \tilde{ {v}}^{F,(t)}_k $ are the $ k^{th} $ DFT-coefficients of $ \tilde{\vec{w}}$ and $ \tilde{ \vec{v}}^{(t)} $, respectively.

\section{Experimental Results}
\label{sec:Experimental Results}

In this section we employ Algorithm \ref{Algorithm:Proposed Method} to adjust standard compression designs to acquisition-rendering systems where the compression is an intermediate stage.
Specifically, we model the source as a high-resolution discrete signal, acquired via linear shift-invariant low-pass filtering and uniform sub-sampling. Then, this acquired signal is available for compression, and after decompression it is linearly rendered back to the source resolution by replicating each decompressed sample (in a uniform pattern matching the source sub-sampling).

\subsubsection{Coding of One Dimensional Signals}
\label{subsec:Experimental Results - 1d signals}

The compression approach here relies on adaptive tree-based segmentation of the signal (for another instance of this prevalent idea see \cite{shukla2005rate}). Specifically, here we compress a one-dimensional signal using an operational rate-distortion optimization determining a binary-tree corresponding to a non-uniform segmentation of the signal, where each segment is represented by a constant value defined by the quantized average-value of the interval (see more details in Appendix D).
We consider the system flow of acquisition-compression-decompression-rendering for an amplitude-modulated chirp source signal. 
Figures \ref{fig:experiments_1d_result_regular} and \ref{fig:experiments_1d_result_proposed} exhibit the system output (i.e., the rendered signal) resulting from the use of a regular compression at 4.71 bpp and from employing our method at 3.69 bpp, respectively. Evidently, our method outperforms the regular approach in terms of PSNR and also in reproducing the chirp signal peaks. Comparing our method to the regular approach at various bit-rates (Fig. \ref{fig:experiments_1d_psnr_rate_curves}) shows significant PSNR gains at medium/high bit-rates.

\subsubsection{Video Coding}
\label{subsec:Experimental Results - video}

We evaluated our method also for adjusting the HEVC coding standard \cite{sullivan2012overview} to a simplified acquisition-rendering system, considering the spatial dimensions of the frames (more details are provided in Appendix E). The PSNR-bitrate curves in Fig. \ref{Fig:Experimental Results - video - PSNR-rate curves} and the visual results (see Appendix E) show the capability of our approach for generically adapting a complicated compression method to a given system structure.

\begin{figure}
	\centering
	{\subfloat['Stockholm' video]{\label{fig:stockholm_psnr_rate_curves}\includegraphics[width=0.20\textwidth]{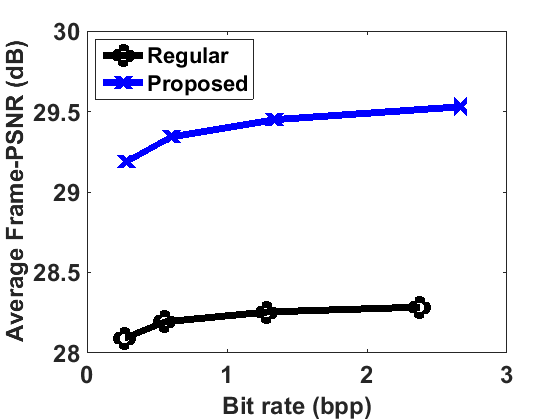}}}~~
	{\subfloat['Shields' video]{\label{fig:shields_psnr_rate_curves}\includegraphics[width=0.20\textwidth]{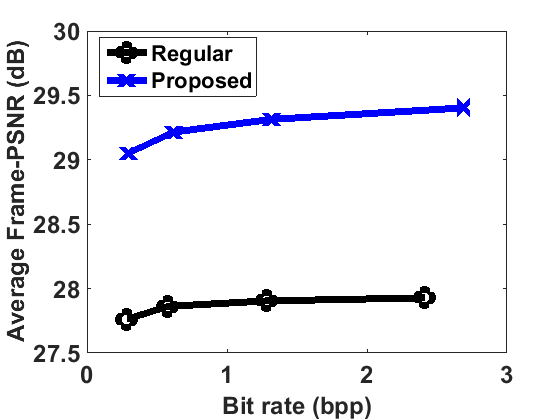}}}
	\caption{HEVC coding of 10 video frames. The curves present the PSNR of the system rendered output with respect to the high-resolution source.} 
	\label{Fig:Experimental Results - video - PSNR-rate curves}
\end{figure}

\section{Conclusion}
\label{sec:Conclusion}
In this paper we considered a system employing lossy compression as an intermediate stage, and proposed a methodology to optimize the system rate-distortion performance from the compression standpoint. 
We presented a generic operational method and explained its main ideas using rate-distortion theory of Gaussian signals. We provided experimental demonstrations of the effectiveness of our approach for coding of 1D signals and video sequences.

\section*{Acknowledgment}
This research was supported in part by the European Research Council under European Union’s Seventh Framework Program, ERC Grant agreement no. 320649, and by the Israel Science Foundation grant no. 2597/16.%
%

\ifCLASSOPTIONcaptionsoff
  \newpage
\fi



\bibliographystyle{IEEEtran}
\bibliography{IEEEabrv,system_aware_compression_conference__refs}
%

%
%

%





\appendices

\section{The Theoretic Settings: Minimal Expected Distortion}
\label{Appendix:theory:subsec:Minimal Expected Distortion}

The minimal distortion was defined in (\ref{eq:suggested distortion  - compression - ideal solution}) for the operational settings. Here we will formulate its expected value, needed for the statistical settings of Section \ref{sec:Theory}. Expecting (\ref{eq:suggested distortion  - compression - ideal solution}), considering $ M = N $ and no noise, gives 
\begin{IEEEeqnarray}{rCl}
	\label{eq:appendix - theory - expected D_0}
	E \left\lbrace D_0 \right\rbrace & = & \frac{1}{N} E \left\lbrace   \left\| { \mtx{A} \left( \mtx{I} - \mtx{P}_{\mtx{B}} \right) \vec{x} } \right\|_2^2   \right\rbrace 
	\nonumber \\ \nonumber
	& = & \frac{1}{N} E \left\lbrace   \left\| { \mtx{F}^{*} \mtx{\Lambda}_{\mtx{A}} \mtx{F} \mtx{F}^{*}\left( \mtx{I} - \mtx{\Lambda}_{\mtx{P}_{\mtx{B}}} \right)\mtx{F} \vec{x} } \right\|_2^2   \right\rbrace 
	\\ \nonumber
	& = & \frac{1}{N} E \left\lbrace   \left\| { \mtx{\Lambda}_{\mtx{A}} \left( \mtx{I} - \mtx{\Lambda}_{\mtx{P}_{\mtx{B}}} \right) \vec{x}^{F} } \right\|_2^2   \right\rbrace 
	\\ \nonumber
	& = & \frac{1}{N} E \left\lbrace  \sum\limits_{k: a_k^F \ne 0 ~,~ b_k^F = 0}   \left\lvert   { a_k^F {x}_k^{F} } \right\rvert ^2   \right\rbrace
	\\ 
	& = & \frac{1}{N} \sum\limits_{k: a_k^F \ne 0 ~,~ b_k^F = 0} {\left\lvert a_k^F \right\rvert ^2 \lambda_k^{\left( \vec{x}\right)} } , 
\end{IEEEeqnarray}
where $ \vec{x}^{F} \triangleq \mtx{F} \vec{x} $. Furthermore, we used the circulant structure of $ \mtx{P}_{\mtx{B}} $ (as it is the projection matrix corresponding to the range of the circulant matrix $ \mtx{B} $) providing the DFT-based diagonalization via $ \mtx{F} \mtx{P}_{\mtx{B}} \mtx{F}^{*} = \mtx{\Lambda}_{\mtx{P}_{\mtx{B}}} $. Moreover, we utilized the formation of the diagonal matrix $ \mtx{\Lambda}_{\mtx{P}_{\mtx{B}}} $ where the $\left(k,k\right) $ entry is 1 for $ k $ corresponding to $ b_k^F \ne 0 $, and otherwise the $ \left(k,k\right) $ entry is 0.

\section{The Theoretic Settings: Equivalence of Optimization Problems (\ref{eq:theory - rate-distortion optimization - basic}) and  (\ref{eq:theory - rate-distortion optimization - pseudoinverse filtered input})}
\label{Appendix:theory:subsec:Equivalence of Optimization Problems 1 and 2}

Let us begin by proving the equivalence between the distortion constraints of optimizations (\ref{eq:theory - rate-distortion optimization - basic}) and (\ref{eq:theory - rate-distortion optimization - pseudoinverse filtered input}). 
We define the matrix $ \mtx{H} \triangleq \mtx{A} \mtx{B}  $ joining the two linear operators of the system. In Section \ref{sec:Theory} we considered $ \mtx{A} $ and $ \mtx{B} $ to be $ N \times N $ circulant matrices and, thus, $ \mtx{H} $ is also a $ N \times N $ circulant matrix. 
Consequently, $ \mtx{H} $ is diagonalized by the $ N\times N $ DFT matrix 
$ \mtx{F} $ via $\mtx{F} \mtx{H} \mtx{F}^* = \mtx{\Lambda}_{\mtx{H}} $, where $ \mtx{\Lambda}_{\mtx{H}} $ is a diagonal matrix formed by the elements $  h^F_k = a^F_k b^F_k $ for $k=0,...,N-1$. 
The pseudoinverse matrix of $ \mtx{H} $ is defined as $\mtx{H}^{+} =  \mtx{F}^* \mtx{\Lambda}_{\mtx{H}}^{+} \mtx{F} $, where $ \mtx{\Lambda}_{\mtx{H}}^{+} $ is the pseudoinverse of $ \mtx{\Lambda}_{\mtx{H}} $, an $ N \times N $ diagonal matrix with the $ k^{th} $ diagonal component
\begin{IEEEeqnarray}{rCl}
\label{eq:appendix - theory - pseudoinverse of Lambda_H - diagonal elements}
h^{F,+}_k = \left\{ {\begin{array}{*{20}{c}}
		{\frac{1}{a^F_k b^F_k}}&{,\text{for~~} k\in \mathcal{K}_{AB} } \\ 
		0&{,\text{for~~} k\notin \mathcal{K}_{AB}. }
	\end{array}} \right.  
\end{IEEEeqnarray} 
where  $ \mathcal{K}_{AB} \nolinebreak \triangleq \nolinebreak \left\lbrace  k : a_k^F \ne 0 ~\text{and}~ b_k^F \ne 0 \right\rbrace $ includes the DFT-domain component indices defining the range of the joint operator $ \mtx{AB} $.

Now, let us develop the distortion expression appearing in the constraint of optimization (\ref{eq:theory - rate-distortion optimization - basic}): 
\begin{IEEEeqnarray}{rCl}
	\label{eq:appendix - theory - equivalence of optimization problems - distortion}
	\left\| { \vec{w}  - \mtx{A} \mtx{B} \vec{v} } \right\|_2^2  & = & \left\| { \vec{w}  - \mtx{H} \vec{v} } \right\|_2^2
	\nonumber \\ 
	 & = & \left\| { \left( \mtx{I} - \mtx{H}\mtx{H}^{+} \right)\vec{w} + \mtx{H}\mtx{H}^{+} \vec{w} - \mtx{H} \vec{v} } \right\|_2^2 
	\nonumber \\  
	& = & \left\| { \left( \mtx{I} - \mtx{H}\mtx{H}^{+} \right)\vec{w} + \mtx{H} \left(  \mtx{H}^{+} \vec{w} - \vec{v} \right) } \right\|_2^2 
	\nonumber \\  
	& = & \left\| {  \left( \mtx{I} - \mtx{H}\mtx{H}^{+} \right) \vec{w} } \right\|_2^2  + \left\| {  \mtx{H} \left( \mtx{H}^{+} \vec{w} -  \vec{v} \right) } \right\|_2^2 \nonumber\\ 
	&& + \left( \mtx{H}^{+} \vec{w} -  \vec{v} \right)^{*} \mtx{H}^{*} \left( \mtx{I} - \mtx{H}\mtx{H}^{+} \right) \vec{w} \nonumber\\ 
	&& + \vec{w}^{*} \left( \mtx{I} - \mtx{H}\mtx{H}^{+} \right)^{*} \mtx{H} \left(  \mtx{H}^{+} \vec{w} - \vec{v} \right)
	\nonumber \\  \nonumber
	& = & \left\| {  \left( \mtx{I} - \mtx{H}\mtx{H}^{+} \right) \vec{w} } \right\|_2^2  + \left\| {  \mtx{H} \left( \mtx{H}^{+} \vec{w} -  \vec{v} \right) } \right\|_2^2  ~~~~ \\
	\end{IEEEeqnarray}
where the last equality readily stems from the relation 
	\begin{IEEEeqnarray}{rCl}
	\label{eq:appendix - theory - equivalence of optimization problems - distortion - auxiliary result}
		\mtx{H}^{*}  \left( \mtx{I} - \mtx{H}\mtx{H}^{+} \right)  = \mtx{0} .
	\end{IEEEeqnarray}
		
We use the DFT-based diagonalization of $ \mtx{H} $ to continue developing the expression of the first term in (\ref{eq:appendix - theory - equivalence of optimization problems - distortion}): 
\begin{IEEEeqnarray}{rCl}
\label{eq:appendix - theory - equivalence of optimization problems - distortion - first part - deterministic}
	\left\| {  \left( \mtx{I} - \mtx{H}\mtx{H}^{+} \right) \vec{w} } \right\|_2^2  & = & \left\| {  \left( \mtx{I} - \mtx{F}^{*}\mtx{\Lambda}_{\mtx{H}} \mtx{F} \mtx{F}^{*}\mtx{\Lambda}_{\mtx{H}}^{+} \mtx{F} \right) \vec{w} } \right\|_2^2 \nonumber \\
	& = & \left\| { \mtx{F}^{*} \left( \mtx{I} - \mtx{\Lambda}_{\mtx{H}} \mtx{\Lambda}_{\mtx{H}}^{+}  \right)\mtx{F} \vec{w} } \right\|_2^2 	\nonumber \\
	& = & \left\| { \left( \mtx{I} - \mtx{\Lambda}_{\mtx{H}} \mtx{\Lambda}_{\mtx{H}}^{+}  \right) \vec{w}^{F}  } \right\|_2^2
	\nonumber \\
	& = & \sum\limits_{k: h_k^F = 0} \left\lvert w_k^F  \right\rvert ^2 = \nonumber \\
	& = &  \sum\limits_{k: a_k^F \ne 0 , b_k^F = 0 } \left\lvert a_k^F x_k^F  \right\rvert ^2 
\end{IEEEeqnarray}
where we used the DFT-domain expression of the $ k^{th} $ component of $ \vec{w}^F $ as $ w_k^F = a_k^F x_k^F  $, reducing to $ w_k^F = 0 $ when $ a_k^F = 0 $. 
Taking the expectation of (\ref{eq:appendix - theory - equivalence of optimization problems - distortion - first part - deterministic}), noting that the source signal $ \vec{x} $ and the noise $ \vec{n} $ are independent, yields 
\begin{IEEEeqnarray}{rCl}
\label{eq:appendix - theory - equivalence of optimization problems - distortion - first part}
	&& E\left\lbrace \left\| {  \left( \mtx{I} - \mtx{H}\mtx{H}^{+} \right) \vec{w} } \right\|_2^2  \right\rbrace =  \sum\limits_{k: a_k^F \ne 0 ~,~ b_k^F = 0} {\left\lvert a_k^F \right\rvert ^2 \lambda_k^{\left( \vec{x}\right)}} .
\end{IEEEeqnarray}
Then, using (\ref{eq:appendix - theory - equivalence of optimization problems - distortion}) and (\ref{eq:appendix - theory - equivalence of optimization problems - distortion - first part}) we get that 
\begin{IEEEeqnarray}{rCl}
	\label{eq:appendix - theory - equivalence of optimization problems - distortion of problem 1 - metric}
	E\left\lbrace \left\| { \vec{w} - \mtx{A} \mtx{B}\vec{v} } \right\|_2^2 \right\rbrace & = & E\left\lbrace \left\| { \mtx{A} \mtx{B} \left( \tilde{\vec{w}} - \vec{v} \right) } \right\|_2^2 \right\rbrace \\ \nonumber && + \sum\limits_{k: a_k^F \ne 0 ~,~ b_k^F = 0} {\left\lvert a_k^F \right\rvert ^2 \lambda_k^{\left( \vec{x}\right)}} , 
\end{IEEEeqnarray} 
where $ \tilde{\vec{w}} = \mtx{B}^{+} \mtx{A}^{+} \vec{w} $.
Eq. (\ref{eq:appendix - theory - expected D_0}) implies 
\begin{IEEEeqnarray}{rCl}
	\label{eq:appendix - theory - equivalence of optimization problems - distortion of problem 1 - ED_0 Reminder}
	N E\left\lbrace D_0 \right\rbrace = \sum\limits_{k: a_k^F \ne 0 ~,~ b_k^F = 0} {\left\lvert a_k^F \right\rvert ^2 \lambda_k^{\left( \vec{x}\right)}} 
\end{IEEEeqnarray} 
that jointly with (\ref{eq:appendix - theory - equivalence of optimization problems - distortion of problem 1 - metric}) yields the transformation of the expected distortion constraint in the optimization (\ref{eq:theory - rate-distortion optimization - basic}), namely, 
\begin{IEEEeqnarray}{rCl}
	\label{eq:appendix - theory - equivalence of optimization problems - distortion of problem 1}
	N  E \left\lbrace D_0 \right\rbrace  \le E\left\lbrace \left\| { \vec{w} - \mtx{A} \mtx{B}\vec{v} } \right\|_2^2 \right\rbrace \le N \left(  E \left\lbrace D_0 \right\rbrace + D \right) ~~
	\end{IEEEeqnarray} 
into 
	\begin{IEEEeqnarray}{rCl}
		\label{eq:appendix - theory - equivalence of optimization problems - distortion of problem 2}
		0 \le E\left\lbrace \left\| { \mtx{A} \mtx{B} \left( \tilde{\vec{w}} - \vec{v} \right) } \right\|_2^2 \right\rbrace \le  N D .
	\end{IEEEeqnarray} 
Since (\ref{eq:appendix - theory - equivalence of optimization problems - distortion of problem 2}) is the distortion constraint in optimization (\ref{eq:theory - rate-distortion optimization - pseudoinverse filtered input}), we proved that the distortion constraints of optimizations (\ref{eq:theory - rate-distortion optimization - basic}) and (\ref{eq:theory - rate-distortion optimization - pseudoinverse filtered input}) are interchangeable.

We continue by showing the equivalence of the mutual information terms appearing in the optimization costs of (\ref{eq:theory - rate-distortion optimization - basic}) and (\ref{eq:theory - rate-distortion optimization - pseudoinverse filtered input}). 
We start with $ I \left( \vec{w}; \mtx{B}\vec{v} \right) $ used in (\ref{eq:theory - rate-distortion optimization - basic}). By the data processing inequality, the compression-input $ \vec{w} $ and its filtered version $ \mtx{A}^{+} {\vec{w}} $ obey 
\begin{IEEEeqnarray}{rCl}
	\label{eq:appendix - theory - equivalence of optimization problems - mutual information - data processing inequality}
	I \left( \vec{w}; \mtx{B}\vec{v} \right) \ge I \left( \mtx{A}^{+} {\vec{w}}; \mtx{B}\vec{v} \right) . 
\end{IEEEeqnarray} 
In the theoretic settings we consider the noiseless case where $ \vec{w} = \mtx{A} \vec{x} $, hence, $ \mtx{A}^{+} {\vec{w}} = \mtx{P}_{\mtx{A}} \vec{x} $. Recall that $ \mtx{P}_{\mtx{A}} $ is the projection matrix of the range of $ \mtx{A} $. Since $ \mtx{A} $ is circulant, $ \mtx{P}_{\mtx{A}} $ is also circulant and diagonalized by the DFT matrix via $ \mtx{F} \mtx{P}_{\mtx{A}} \mtx{F}^{*} = \mtx{\Lambda}_{\mtx{P}_{\mtx{A}}} $ where $\mtx{\Lambda}_{\mtx{P}_{\mtx{A}}} = \mtx{\Lambda}_{{\mtx{A}}} \mtx{\Lambda}_{{\mtx{A}}}^{+} $.
Accordingly, we can write 
\begin{IEEEeqnarray}{rCl}
	\label{eq:appendix - theory - equivalence of optimization problems - mutual information - data processing inequality - w via w_tilde}
	\mtx{A} \mtx{A}^{+} {\vec{w}} = \mtx{A} \mtx{P}_{\mtx{A}} \vec{x} = \mtx{A} \vec{x} = \vec{w}
\end{IEEEeqnarray}
exhibiting $\vec{w}$ as a processing of $ \mtx{A}^{+} {\vec{w}}  $, thus, by the data processing inequality we get 
\begin{IEEEeqnarray}{rCl}
	\label{eq:appendix - theory - equivalence of optimization problems - mutual information - data processing inequality - second direction}
	I \left( \vec{w}; \mtx{B}\vec{v} \right) \le I \left( \mtx{A}^{+} {\vec{w}}; \mtx{B}\vec{v} \right)
\end{IEEEeqnarray} 
that together with (\ref{eq:appendix - theory - equivalence of optimization problems - mutual information - data processing inequality}) implies 
\begin{IEEEeqnarray}{rCl}
	\label{eq:appendix - theory - equivalence of optimization problems - mutual information - data processing inequality - with equality}
	I \left( \vec{w}; \mtx{B}\vec{v} \right) = I \left( \mtx{A}^{+} {\vec{w}}; \mtx{B}\vec{v} \right) .
\end{IEEEeqnarray} 
The above showed that since $ \vec{w} $ is in the range of $ \mtx{A} $, its processing via $ \mtx{A}^{+} $ is invertible and, thus, the mutual information is preserved under such processing of $ \vec{w} $. 

Similar to the above arguments, we can show that since the system output $ \mtx{B}\vec{v} $ is in the range of $ \mtx{B} $, its processing via $ \mtx{B}^{+} $ is invertible and, therefore, the mutual information is preserved under such filtering of $ \mtx{B}\vec{v} $: 
\begin{IEEEeqnarray}{rCl}
	\label{eq:appendix - theory - equivalence of optimization problems - mutual information - data processing inequality - Bv - with equality}
	I \left( \mtx{A}^{+} {\vec{w}}; \mtx{B}\vec{v} \right) & = & I \left( \mtx{A}^{+} {\vec{w}}; \mtx{B}^{+} \mtx{B}\vec{v} \right) \nonumber \\ 
	& = & I \left( \mtx{A}^{+} {\vec{w}}; \mtx{P}_{\mtx{B}} \vec{v} \right) . 
\end{IEEEeqnarray}
	
Consider the decomposition of $ \mtx{A}^{+} {\vec{w}} $ based on its projections on $ \mtx{B} $'s range and nullspace, namely, 
\begin{IEEEeqnarray}{rCl}
	\label{eq:appendix - theory - equivalence of optimization problems - mutual information - deomposition of A^+ w}
	\mtx{A}^{+} {\vec{w}} =  \mtx{P}_{\mtx{B}} \mtx{A}^{+} {\vec{w}} + \left( \mtx{I} - \mtx{P}_{\mtx{B}} \right) \mtx{A}^{+} {\vec{w}} .
\end{IEEEeqnarray}
Observe that 
\begin{IEEEeqnarray}{rCl}
	\label{eq:appendix - theory - equivalence of optimization problems - mutual information - deomposition of A^+ w - cross covariance - preliminary -  1}
	\left( \mtx{I} - \mtx{P}_{\mtx{B}} \right) \mtx{A}^{+} {\vec{w}} & = & \mtx{F}^{*}\left( \mtx{I} - \mtx{\Lambda}_{\mtx{P}_{\mtx{B}}} \right)\mtx{F} \mtx{F}^{*} \mtx{\Lambda}_{\mtx{A}}^{+} \mtx{F} {\vec{w}} \nonumber \\ 
	& = & \mtx{F}^{*}\left( \mtx{I} - \mtx{\Lambda}_{\mtx{P}_{\mtx{B}}} \right) \mtx{\Lambda}_{\mtx{A}}^{+} {\vec{w}^{F}}
\end{IEEEeqnarray}
and the second random variable considered by the mutual information in (\ref{eq:appendix - theory - equivalence of optimization problems - mutual information - data processing inequality - Bv - with equality}) is 
\begin{IEEEeqnarray}{rCl}
	\label{eq:appendix - theory - equivalence of optimization problems - mutual information - deomposition of A^+ w - cross covariance - preliminary - 2}
	\mtx{P}_{\mtx{B}} \vec{v} & = & \mtx{F}^{*} \mtx{\Lambda}_{\mtx{P}_{\mtx{B}}} \mtx{F}  {\vec{w}} \nonumber \\ 
	& = & \mtx{F}^{*} \mtx{\Lambda}_{\mtx{P}_{\mtx{B}}}  {\vec{w}}^{F}
\end{IEEEeqnarray}
where $ {\vec{w}}^{F} $ is the DFT-domain representation of ${\vec{w}}$. 
Note that the diagonal projection matrices $ \mtx{\Lambda}_{\mtx{P}_{\mtx{B}}} \triangleq \mtx{\Lambda}_{\mtx{B}} \mtx{\Lambda}_{\mtx{B}}^{+} $ and $ \mtx{I} - \mtx{\Lambda}_{\mtx{P}_{\mtx{B}}} $ appearing in (\ref{eq:appendix - theory - equivalence of optimization problems - mutual information - deomposition of A^+ w - cross covariance - preliminary - 2}) and (\ref{eq:appendix - theory - equivalence of optimization problems - mutual information - deomposition of A^+ w - cross covariance - preliminary -  1}), respectively, imply that the DFT-domain representations of $ \left( \mtx{I} - \mtx{P}_{\mtx{B}} \right) \mtx{A}^{+} {\vec{w}}  $ and $ \mtx{P}_{\mtx{B}} \vec{v} $ have complementary non-overlapping supports (defined by the component indices where the value may be nonzero). 
Let us develop the mutual information in (\ref{eq:appendix - theory - equivalence of optimization problems - mutual information - data processing inequality - Bv - with equality}) to leverage the mentioned structures and the decomposition (\ref{eq:appendix - theory - equivalence of optimization problems - mutual information - deomposition of A^+ w}): 
\begin{IEEEeqnarray}{rCl}	
	&&I \left( \mtx{A}^{+} {\vec{w}}; \mtx{P}_{\mtx{B}} \vec{v} \right) =
	\nonumber \\
	\label{eq:appendix - theory - equivalence of optimization problems - mutual information - data processing inequality - Bv - using independence}
	&&= I \left( \mtx{F} \mtx{A}^{+} {\vec{w}}; \mtx{F} \mtx{P}_{\mtx{B}} \vec{v} \right)
	\\
	\label{eq:appendix - theory - equivalence of optimization problems - mutual information - data processing inequality - Bv - using independence - 2}
	&&= I \left( \mtx{\Lambda}_{\mtx{A}}^{+} {\vec{w}^{F}}; \mtx{\Lambda}_{\mtx{P}_{\mtx{B}}} \vec{v}^{F} \right)
	\\
	\label{eq:appendix - theory - equivalence of optimization problems - mutual information - data processing inequality - Bv - using independence - 3}
	&&\ge  \sum\limits_{k: b_k^F \ne 0} {I \left( a_k^{F,+} {{w}_k^{F}}; {v}_k^{F} \right)} + \sum\limits_{k: b_k^F = 0} {I \left( a_k^{F,+} {{w}_k^{F}}; 0 \right)}
	 \\ 
	 \label{eq:appendix - theory - equivalence of optimization problems - mutual information - data processing inequality - Bv - using independence - 4}
	&&= \sum\limits_{k: b_k^F \ne 0} {I \left( a_k^{F,+} {{w}_k^{F}}; {v}_k^{F} \right)}
\end{IEEEeqnarray}
where (\ref{eq:appendix - theory - equivalence of optimization problems - mutual information - data processing inequality - Bv - using independence}) is due to the preservation of mutual information under a unitary transformation. The lower bound in (\ref{eq:appendix - theory - equivalence of optimization problems - mutual information - data processing inequality - Bv - using independence - 3}) emerges from the independence of the DFT-domain variables $ {{w}_k^{F}} $, $ k=0,...,N-1 $, since they originate in the cyclo-stationary Gaussian signal $ \vec{x} $ processed by the circulant matrix $ \mtx{A} $. 
In Appendix \ref{Appendix:theory:subsec:Equivalence of Optimization Problems 2 and 3} we exhibit a backward channel construction based on independent components $ {v}_k^{F} $ in the DFT domain, therefore, satisfying the inequality (\ref{eq:appendix - theory - equivalence of optimization problems - mutual information - data processing inequality - Bv - using independence - 3}) with equality. Using (\ref{eq:appendix - theory - equivalence of optimization problems - mutual information - data processing inequality - Bv - using independence - 4}) and pending on the construction shown in Appendix \ref{Appendix:theory:subsec:Equivalence of Optimization Problems 2 and 3}, we state that 
\begin{IEEEeqnarray}{rCl}
	\label{eq:appendix - theory - equivalence of optimization problems - mutual information - data processing inequality - Bv - using independence - conclude}
	I \left( \mtx{A}^{+} {\vec{w}}; \mtx{P}_{\mtx{B}} \vec{v} \right) & = & \sum\limits_{k: b_k^F \ne 0} {I \left( a_k^{F,+} {{w}_k^{F}}; {v}_k^{F}  \right) }
	\nonumber \\
	& = & I \left( \mtx{P}_{\mtx{B}} \mtx{A}^{+} {\vec{w}}; \mtx{P}_{\mtx{B}} \vec{v} \right)
\end{IEEEeqnarray}


Note that $ \mtx{P}_{\mtx{B}} \mtx{A}^{+} {\vec{w}} $ is in the range of $ \mtx{A} $, hence, using similar arguments to those given in the last paragraph, one can show that 
\begin{IEEEeqnarray}{rCl}
	\label{eq:appendix - theory - equivalence of optimization problems - mutual information - data processing inequality - Bv - using independence - }
	I \left( \mtx{P}_{\mtx{B}} \mtx{A}^{+} {\vec{w}}; \mtx{P}_{\mtx{B}} \vec{v} \right) = I \left( \mtx{P}_{\mtx{B}} \mtx{A}^{+} {\vec{w}}; \mtx{P}_{\mtx{A}}\mtx{P}_{\mtx{B}} \vec{v} \right) 
\end{IEEEeqnarray}
Furthermore, since $ \mtx{P}_{\mtx{B}} \mtx{A}^{+} {\vec{w}} $ belongs also to the range of $ \mtx{B} $, its processing via $ \mtx{B}^{+} $ is invertible, thus, 
\begin{IEEEeqnarray}{rCl}
	\label{eq:appendix - theory - equivalence of optimization problems - mutual information - data processing inequality - Bv - using invertible processing}
	I \left( \mtx{P}_{\mtx{B}} \mtx{A}^{+} {\vec{w}}; \mtx{P}_{\mtx{A}}\mtx{P}_{\mtx{B}} \vec{v} \right) = I \left( \mtx{B}^{+} \mtx{A}^{+} {\vec{w}}; \mtx{P}_{\mtx{A}}\mtx{P}_{\mtx{B}} \vec{v} \right) .
\end{IEEEeqnarray}
Due to the circulant structure of $ \mtx{P}_{\mtx{A}} $ and $ \mtx{P}_{\mtx{B}} $, they commute and, therefore, 
\begin{IEEEeqnarray}{rCl}
	\label{eq:appendix - theory - equivalence of optimization problems - mutual information - data processing inequality - Bv - commute}
	I \left( \mtx{P}_{\mtx{B}} \mtx{A}^{+} {\vec{w}}; \mtx{P}_{\mtx{A}}\mtx{P}_{\mtx{B}} \vec{v} \right) = I \left( \mtx{B}^{+} \mtx{A}^{+} {\vec{w}}; \mtx{P}_{\mtx{B}} \mtx{P}_{\mtx{A}} \vec{v} \right) ,
\end{IEEEeqnarray}
that is the mutual information in the cost of problem (\ref{eq:theory - rate-distortion optimization - pseudoinverse filtered input}). 
To conclude, we showed the equivalence of the costs (based on the backward-channel construction presented next in Appendix \ref{Appendix:theory:subsec:Equivalence of Optimization Problems 2 and 3}) and the constraints of problems (\ref{eq:theory - rate-distortion optimization - basic}) and (\ref{eq:theory - rate-distortion optimization - pseudoinverse filtered input}), therefore, these optimization problems are interchangeable.

\section{The Theoretic Settings: Equivalence of Optimization Problems (\ref{eq:theory - rate-distortion optimization - pseudoinverse filtered input}) and (\ref{eq:theoretic Gaussian analysis - DFT-domain - Gaussian distortion allocation - general})}
\label{Appendix:theory:subsec:Equivalence of Optimization Problems 2 and 3}

Since $ \mtx{A} $ and $ \mtx{B} $ are circulant matrices, the expected distortion $ E\left\lbrace \left\| { \mtx{A} \mtx{B} \left( \tilde{\vec{w}} - \vec{v} \right) } \right\|_2^2 \right\rbrace $, appearing in the constraint of (\ref{eq:theory - rate-distortion optimization - pseudoinverse filtered input}), has the following additively-separable form in the DFT domain 
\begin{IEEEeqnarray}{rCl}
	\label{eq:appendix - theory - equivalence of optimization problems - distortion - separable decomposition}
	E\left\lbrace \left\| { \mtx{A} \mtx{B} \left( \tilde{\vec{w}} - \vec{v} \right) } \right\|_2^2 \right\rbrace & = & E\left\lbrace \left\| { \mtx{\Lambda}_{\mtx{A}} \mtx{\Lambda}_{\mtx{B}} \left( \tilde{\vec{w}}^F - \vec{v}^F \right) } \right\|_2^2 \right\rbrace ~~~~~~
	\nonumber \\ 
	& = & \sum\limits_{k=0}^{N-1} { \left| a^F_k b^F_k \right|^2 E\left\lbrace \left| {  \tilde{w}^F_k - v^F_k } \right|^2 \right\rbrace } .~~~~~~
\end{IEEEeqnarray} 
This expected distortion formulation motivates us to address the entire rate-distortion optimization in the DFT domain and with respect to a pseudoinverse filtered version of the input $ \vec{w} $. For this purpose we will treat next the optimization cost of the problem given in (\ref{eq:theory - rate-distortion optimization - pseudoinverse filtered input}). 

Since $ \tilde{\vec{w}} \triangleq \mtx{B}^{+} \mtx{A}^{+} {\vec{w}} $ is a cyclo-stationary Gaussian signal, we apply the following familiar lower bounds:
	\begin{IEEEeqnarray}{rCl}
		\label{eq:appendix - theory - equivalence of optimization problems - mutual information - Fourier}
		I \left( \tilde{\vec{w}}; \mtx{P}_{\mtx{B}} \mtx{P}_{\mtx{A}}\vec{v} \right) & = & I \left( \mtx{F} \tilde{\vec{w}}; \mtx{F} \mtx{P}_{\mtx{B}} \mtx{P}_{\mtx{A}}\vec{v} \right)
		\\ 
		\label{eq:appendix - theory - equivalence of optimization problems - mutual information - Fourier - specific}
		& = & I \left( \tilde{\vec{w}}^F; \mtx{\Lambda}_{\mtx{P}_{\mtx{B}}} \mtx{\Lambda}_{\mtx{P}_{\mtx{A}}}\vec{v}^F \right)
		\\ 
		\label{eq:appendix - theory - equivalence of optimization problems - mutual information - Fourier - separability}
		& \ge & \sum_{k\in\mathcal{K}_{AB}} { I \left( \tilde{{w}}^F_k; {v}^F_k \right) }
	\end{IEEEeqnarray} 
where (\ref{eq:appendix - theory - equivalence of optimization problems - mutual information - Fourier}) is due to the invariance of mutual information under a unitary transformation, (\ref{eq:appendix - theory - equivalence of optimization problems - mutual information - Fourier - specific}) exhibits the diagonal forms of the projection matrices, and the bound (\ref{eq:appendix - theory - equivalence of optimization problems - mutual information - Fourier - separability}) is due to the independence of the DFT coefficients $ \left\lbrace \tilde{{w}}^F_k \right\rbrace_{k=0}^{N-1} $ stemming from cyclo-stationary Gaussian characteristics of $ \tilde{\vec{w}} $. Moreover, note that (\ref{eq:appendix - theory - equivalence of optimization problems - mutual information - Fourier - separability}) refers only to mutual information of DFT components belonging to the range of $ \mtx{A} \mtx{B} $, this is due to (\ref{eq:appendix - theory - equivalence of optimization problems - mutual information - Fourier - specific}) where $ \vec{v}^F $ components corresponding to the nullspace of $ \mtx{A} \mtx{B} $ are zeroed (hence, they yield zero mutual information and zero rate).
The next lower bound emerges from the definition of the rate-distortion function for each of the components, i.e., 
	\begin{IEEEeqnarray}{rCl}
		\label{eq:appendix - theory - equivalence of optimization problems - mutual information - Gaussian rate-distortion}
		\sum_{k\in\mathcal{K}_{AB}} { I \left( \tilde{{w}}^F_k; {v}^F_k \right) } & \ge & \sum_{k\in\mathcal{K}_{AB}} { R_k \left( D_k \right) }
		\\ 
		\label{eq:appendix - theory - equivalence of optimization problems - mutual information - Gaussian rate-distortion - explicit}
		& = & \sum\limits_{k\in\mathcal{K}_{AB}} { \left[  \frac{1}{2} \log \left( { \frac{ \lambda^{\left(\tilde{\vec{w}}\right)}_k   }{ D_k } } \right) \right]_+ }
	\end{IEEEeqnarray} 
where the last equality relies on the rate-distortion function formulation for a scalar Gaussian source. The $ k^{th} $ variable here is $ \tilde{{w}}^F_k $, having the a variance denoted as $ \lambda^{\left(\tilde{\vec{w}}\right)}_k $. Here, the $ k^{th} $-component rate corresponds to an expected squared-error distortion denoted as $ D_k \triangleq E\left\lbrace \left| {  \tilde{w}^F_k - v^F_k } \right|^2 \right\rbrace $ .

The mutual-information lower bound in (\ref{eq:appendix - theory - equivalence of optimization problems - mutual information - Gaussian rate-distortion - explicit}) is further minimized under the total distortion constraint that expresses the weights introduced in (\ref{eq:appendix - theory - equivalence of optimization problems - distortion - separable decomposition}) for each of the DFT components, i.e., the distortion-allocation optimization is 
\begin{IEEEeqnarray}{rCl}
	\label{eq:appendix - theory - equivalence of optimization problems - distortion allocation - 2}
	\begin{aligned}
		& \underset{\left\lbrace D_k \right\rbrace_{k\in\mathcal{K}_{AB}}}{\text{min}}
		& & \sum\limits_{k\in\mathcal{K}_{AB}} { \frac{1}{2} \log \left( { \frac{ \lambda^{\left(\tilde{\vec{w}}\right)}_k   }{ D_k } } \right)  } \\
		& \text{s.t.}
		& & \sum\limits_{k\in\mathcal{K}_{AB}} { \left| a^F_k b^F_k\right|^2 D_k } \le N D ~~~
		\\
		& & & 0 \le D_k \le \lambda^{\left(\tilde{\vec{w}}\right)}_k ~~~,~k\in\mathcal{K}_{AB}.
	\end{aligned}
\end{IEEEeqnarray}
where the operator $ \left[ \cdot \right]_+ $ in (\ref{eq:appendix - theory - equivalence of optimization problems - mutual information - Gaussian rate-distortion - explicit}) was replaced with componentwise constraints.
The optimization (\ref{eq:appendix - theory - equivalence of optimization problems - distortion allocation - 2}) is solved using Lagrangian optimization and the KKT conditions. We denote here the optimal distortions as $ \left\lbrace \hat{D}_k \right\rbrace_{k = 0}^{N-1} $, where for $ k \notin \mathcal{K}_{AB} $ we set $ \hat{D}_k = 0 $, and use them next for showing the achievability of the mutual-information lower bound.

We define $ \mtx{\Lambda}_{\hat{D}}  $ as the $ N\times N $ diagonal matrix with $ \hat{D}_k $ as the $ k^{th} $ diagonal value. Also recall that $ \mtx{H} \triangleq \mtx{A} \mtx{B} $ and the related definitions given above. Consider the following construction for a backward channel producing $ \tilde{\vec{w}} $ from $ \vec{v} $. 
Let 
\begin{IEEEeqnarray}{rCl}
\label{eq:appendix - proofs - equivalence of problem 1 and 2 - backward channel construction - x }
{\vec{v}} & \sim & \mathcal{N} \left( \vec{0}, \mtx{H}^{+} \mtx{R}_{\vec{w}} \mtx{H}^{+*} - \mtx{H}^{+} \mtx{F}^{*} \mtx{\Lambda}_{\hat{D}} \mtx{F} \mtx{H}^{+*} \right) 
\\ 
\label{eq:appendix - proofs - equivalence of problem 1 and 2 - backward channel construction - z }
\vec{z} & \sim & \mathcal{N} \left( \vec{0}, \mtx{H}^{+} \mtx{F}^{*} \mtx{\Lambda}_{\hat{D}} \mtx{F} \mtx{H}^{+*} \right) 
\end{IEEEeqnarray} 
be two independent random vectors, constructing $ \tilde{\vec{w}} $ via 
\begin{IEEEeqnarray}{rCl}
\label{eq:appendix - proofs - equivalence of problem 1 and 2 - backward channel construction - y_tilde as a sum }
\tilde{\vec{w}} = {\vec{v}} + \vec{z} ,
\end{IEEEeqnarray} 
hence, $ \tilde{\vec{w}}  \sim \mathcal{N} \left( \vec{0}, \mtx{H}^{+} \mtx{R}_{\vec{w}} \mtx{H}^{+*} \right) $, agreeing with $ \tilde{\vec{w}} \nolinebreak = \nolinebreak \mtx{H}^{+} \vec{w} $ where $\vec{w} \sim \mathcal{N} \left( \vec{0}, \mtx{R}_{\vec{w}} \right)$. 
Additionally, the construction (\ref{eq:appendix - proofs - equivalence of problem 1 and 2 - backward channel construction - x })-(\ref{eq:appendix - proofs - equivalence of problem 1 and 2 - backward channel construction - y_tilde as a sum }) leads to 
\begin{IEEEeqnarray}{rCl}
\label{eq:appendix - proofs - equivalence of problem 1 and 2 - backward channel construction - distortion constraint satisfies}
E\left\lbrace \left\| \mtx{H} \left(  { {\tilde{\vec{w}}}  -  {\vec{v}} } \right) \right\|_2^2  \right\rbrace & = & E\left\lbrace \left\| \mtx{H} \vec{z} \right\|_2^2  \right\rbrace \nonumber
\\
& = & E\left\lbrace \vec{z}^{*}  \mtx{H}^{*}  \mtx{H} \vec{z} \right\rbrace \nonumber\\
& = & E\left\lbrace Trace\left\lbrace \vec{z}^{*}  \mtx{H}^{*}  \mtx{H} \vec{z} \right\rbrace \right\rbrace \nonumber\\
& = & E\left\lbrace Trace\left\lbrace \mtx{H} \vec{z} \vec{z}^{*}  \mtx{H}^{*}   \right\rbrace \right\rbrace \nonumber\\
& = & Trace\left\lbrace \mtx{H} \mtx{R}_{\vec{z}} \mtx{H}^{*}   \right\rbrace 
\nonumber\\
& = & Trace\left\lbrace \mtx{H} \mtx{H}^{+} \mtx{F}^{*} \mtx{\Lambda}_{\hat{D}} \mtx{F} \mtx{H}^{+*} \mtx{H}^{*}   \right\rbrace 
\nonumber\\
& = & Trace\left\lbrace \mtx{F}^{*} \mtx{\Lambda}_{\mtx{H}} \mtx{\Lambda}_{\mtx{H}}^{+}  \mtx{\Lambda}_{\hat{D}} \mtx{\Lambda}_{\mtx{H}}^{+*} \mtx{\Lambda}_{\mtx{H}}^{*} \mtx{F} \right\rbrace 
\nonumber\\
& = & Trace\left\lbrace \mtx{F}^{*}  \mtx{\Lambda}_{\hat{D}} \mtx{F} \right\rbrace 
\nonumber\\
& = & Trace\left\lbrace \mtx{\Lambda}_{\hat{D}}  \right\rbrace 
\nonumber\\
& = &  N D 
\end{IEEEeqnarray} 
that reaches the maximal allowed distortion in (\ref{eq:appendix - theory - equivalence of optimization problems - distortion allocation - 2}).
We used here the relation $ \mtx{\Lambda}_{\mtx{H}} \mtx{\Lambda}_{\mtx{H}}^{+}  \mtx{\Lambda}_{\hat{D}} \mtx{\Lambda}_{\mtx{H}}^{+*} \mtx{\Lambda}_{\mtx{H}}^{*} = \mtx{\Lambda}_{\hat{D}} $, emerging from the fact that $ \hat{D}_k = 0 $ for components in the nullspace of $ \mtx{H} $.
This construction fulfills (\ref{eq:appendix - theory - equivalence of optimization problems - mutual information - data processing inequality - Bv - using independence - 3}), (\ref{eq:appendix - theory - equivalence of optimization problems - mutual information - Fourier - separability}) and (\ref{eq:appendix - theory - equivalence of optimization problems - mutual information - Gaussian rate-distortion}) with equality, thus, proves the equivalence of the optimization problems (\ref{eq:theory - rate-distortion optimization - basic}), (\ref{eq:theory - rate-distortion optimization - pseudoinverse filtered input}), and (\ref{eq:theoretic Gaussian analysis - DFT-domain - Gaussian distortion allocation - general}).

\section{Additional Details on the Experiments for Coding of One-Dimensional Signals}
\label{Appendix:experiments:1D signals}

\subsubsection{The Tree-based Coding Method for One-Dimensional Signals}
\label{Appendix:experiments:1D signals - The Tree-based Coding Method for One-Dimensional Signals}
~

We consider a coding procedure that relies on a nonuniform segmentation of the signal based on a binary tree structure. The method presented here is influenced by the general framework given in \cite{chou1989optimal} for optimizing tree-structures, and by the rate-distortion Lagrangian optimization in \cite{shoham1988efficient,ortega1998rate}. 

We consider the coding of a $ M $-length vector $ \vec{w} \in \mathbb{R}^{M} $, where $ M=2^{d_0} $ for some positive integer $ d_0 $. 
The procedure starts with a full $ d $-depth binary-tree ($ d \le d_0 $), which is the initial tree, describing a uniform partitioning of the vector components $ w_k $, $k=0,...,M-1$, into $ 2^d $ sub-vectors of $ M\cdot 2^{-d} $ length. The segmentation of the vector is represented by the leaves of the binary tree: the sub-vector location and length are determined by the leaf place in the tree, in particular, the sub-vector length is defined by the tree-level that the leaf belongs to.
The examined nonuniform segmentations are induced by all the trees obtained by repeatedly pruning neighboring-leaves having the same parent node. The initial $ d $-depth full-tree together with all its pruned subtrees form the set of relevant trees, denoted here as $ \mathcal{T}_d $.

The leaves of a tree $ T \in \mathcal{T}_d $ form a set denoted as $ L(T) $, where the number of leaves is referred to as $ |L(T)| $. Accordingly, the tree $ T $ represents a (possibly) nonuniform partitioning of the $ M $-length vector into $ |L(T)| $ segments. A leaf $ l\nolinebreak\in\nolinebreak L(T) $ resides in the $ h(l) $ level of the tree and corresponds to the indexing interval $ \left[ a^{left}_{(l)} ,..., a^{right}_{(l)}  \right] $ of length $\Delta\left(l\right)\nolinebreak=\nolinebreak M\cdot 2^{-h(l)}$. 
A segment, corresponding to the leaf $ l\in L(T) $, is represented by its average value
\begin{IEEEeqnarray}{rCl}
\label{eq:tree-structured sampling - optimal sample}
{\hat{w}_{(l)}} = \frac{1}{\Delta\left(l\right)} \mathop \sum_{k = a^{left}_{(l)}}^{a^{right}_{(l)}} w_k 
\end{IEEEeqnarray}
that is further uniformly quantized using $q_b = 8$ bits. The quantized sample corresponding to the $ l^{th} $ leaf (segment) is denoted as $ {\hat{w}_{(l)}^Q} $.
This coding structure leads to reconstruction squared-error induced by the tree $ T \in \mathcal{T}_d $ and calculated based on its leaves, $L(T)$, via
\begin{IEEEeqnarray}{rCl}
\label{eq:tree-structured sampling - sampling MSE for a tree}
\mathcal{E}^2 \left( T \right) = \mathop\sum_{l\in L(T)} {  \mathop \sum \limits_{k = a^{left}_{(l)}}^{a^{right}_{(l)}}  {\left(  w_k - {\hat{w}_{(l)}^Q}  \right) ^2 }    } .
\end{IEEEeqnarray}

For a given signal $ \vec{w} \in \mathbb{R}^{M} $ and a budget of $ \rho $ bits, one can formulate the optimization of a tree-structured nonuniform coding as
\begin{IEEEeqnarray}{rCl}
\label{eq:tree-structured sampling - constrained optimization}
\begin{aligned}
& \underset{T\in\mathcal{T}_d}{\text{minimize}}
& & \mathcal{E}^2 \left( T \right) \\
& \text{subject to}
& & q_b|L(T)| = \rho , 
\end{aligned}
\end{IEEEeqnarray}
namely, the optimization searches for the tree associated with a bit-cost of $ \rho $ bits that provides minimal reconstruction squared-error. The unconstrained Lagrangian form of (\ref{eq:tree-structured sampling - constrained optimization}) is 
\begin{IEEEeqnarray}{rCl}
\label{eq:tree-structured sampling - unconstrained Lagrangian optimization}
\underset{T\in\mathcal{T}_d}{\min} \left\lbrace \mathcal{E}^2 \left( T \right) + \nu \left( q_b|L(T)|\right)  \right\rbrace, 
\end{IEEEeqnarray}
where $ \nu \ge 0 $ is a Lagrange multiplier that corresponds to $q_b|L(T)| = \rho$. However, it should be noted that due to the discrete nature of the problem such $ \nu $ does not necessarily exist for any $\rho$ value (see details, e.g., in \cite{chou1989optimal,everett1963generalized}).
The problem (\ref{eq:tree-structured sampling - unconstrained Lagrangian optimization}) can also be written as
\begin{IEEEeqnarray}{rCl}
\label{eq:tree-structured sampling - unconstrained Lagrangian optimization - explicit}
\underset{T\in\mathcal{T}_d}{\min} \left\lbrace \mathop\sum_{l\in L(T)} {  \mathop \sum \limits_{k = a^{left}_{(l)}}^{a^{right}_{(l)}}  {\left(  w_k - {\hat{w}_{(l)}^Q}  \right) ^2 } } + \nu q_b |L(T)| \right\rbrace.
\end{IEEEeqnarray}
Importantly, since the representation intervals do not overlap, the contribution of a leaf, $ l \in L(T) $, to the Lagrangian cost is 
\begin{IEEEeqnarray}{rCl}
\label{eq:tree-structured sampling - leaf Lagrangian cost}
C\left(l\right) = \mathop \sum \limits_{k = a^{left}_{(l)}}^{a^{right}_{(l)}}  {\left(  w_k - {\hat{w}_{(l)}^Q}  \right) ^2 }  + \nu q_b .
\end{IEEEeqnarray}	

The discrete optimization problem (\ref{eq:tree-structured sampling - unconstrained Lagrangian optimization - explicit}) of optimizing the tree for a given signal and a Lagrange multiplier $ \nu $ is practically addressed as follows. 
Start from the full $d$-depth tree and determine the corresponding segments and their quantized samples, squared errors, and contributions to the Lagrangian cost (\ref{eq:tree-structured sampling - leaf Lagrangian cost}). Go through the tree levels from bottom and up, in each tree level find the pairs of neighboring leaves having the same parent node and evaluate the pruning condition: if 
\begin{IEEEeqnarray}{rCl}
\label{eq:tree-structured sampling - leaf pruning condition}
C\left(\text{left child}\right) + C\left(\text{right child}\right) > C\left(\text{parent}\right)
\end{IEEEeqnarray}	
is true, then prune the two leaves -- implying that two segments are merged to form a single sub-vector of double length (thus, the total bit-cost is reduced by $q_b$). If the condition (\ref{eq:tree-structured sampling - leaf pruning condition}) is false, then the two leaves (and the associated representation segments) are kept.
This evaluation is continued until reaching a level where no pruning is done, or when arriving to the root of the tree.

\subsubsection{The Experiment Settings used for Adjusting Compression of One-Dimensional Signals to an Acquisition-Rendering System}
\label{Appendix:experiments:1D signals - Adjusting to System Experiment}
~

The considered source signal $ \vec{x} $ is an amplitude-modulated chirp (see the blue curves in Figs. \ref{fig:experiments_1d_result_regular}-\ref{fig:experiments_1d_result_proposed}), defined by sampling its mathematical formulation using 1024 samples uniformly spread in the "time" interval $ [0,1) $. Note that the chirp signal values are in the range $ [0,1] $, a property used in the PSNR computation shown in the result evaluation in Fig. \ref{fig:experiments_1d_psnr_rate_curves}.

The acquisition is modeled here by a low-pass filtering applied using a convolution with a Gaussian kernel (standard deviation 15) and support of 15 samples, followed by sub-sampling in a factor of 4, and an additive white Gaussian noise with standard deviation 0.001. This procedure results with the 256-samples signal $ \vec{w} $ that is given to the compression. 
After decompression the rendering operator is applied by replicating each sample of $ \vec{v} $ four times such that the piecewise-constant signal $ \vec{y} $ is formed, having a size of 1024 samples.

We tested two compression strategies to employ in the acquisition-rendering system. The first is a regular compression using the tree-based procedure, described in the former subsection, applied based on some Lagrange multiplier $ \nu $. The experiment for this regular flow was repeated for various values of the Lagrange multiplier $ \nu $ to produce the PSNR-bitrate curve in Fig. \ref{fig:experiments_1d_psnr_rate_curves}. 
The second approach relied on the implementation of the proposed method (Algorithm \ref{Algorithm:Proposed Method}) as the compression stage that leverages the standard tree-based coding described in the former subsection. This approach was evaluated for a range of  $ \nu $ values to obtain the corresponding PSNR-bitrate curve in Fig. \ref{fig:experiments_1d_psnr_rate_curves}. The implementation of our iterative method (Algorithm \ref{Algorithm:Proposed Method}) run a maximum of 40 iterations or until convergence is detected.

\section{Additional Details on the Experiments for Video Coding}
\label{Appendix:experiments:video}

The source signal $ \vec{x} $ is a sequence of 10 frames, each of $ 480\times 480 $ pixels.
The acquisition is modeled here by a low-pass filtering carried out by a convolution with a two-dimensional Gaussian kernel (standard deviation 1) and support of $ 5\times 5 $ pixels, followed by horizontal and vertical sub-sampling in a factor of 2, and an additive white Gaussian noise with standard deviation 0.001. This procedure results with the 10-frame sequence $ \vec{w} $ with a frame size of $ 240 \times 240 $ pixels. 
The rendering applied after decompression is simply done by replicating each pixel of $ \vec{v} $ in a $ 2\times 2 $ pixels square such that the rendered signal $ \vec{y} $ has frames of $ 480\times 480 $ pixels having spatial piecewise-constant form.

We evaluated two compression approaches to apply in the acquisition-rendering system. One, is to employ a regular compression using the HEVC video coding standard (using its reference software, version HM 15.0, available at http://hevc.hhi.fraunhofer.de/). The experiment for this regular approach was repeated for various values of the quality parameter of the HEVC to produce the PSNR-bitrate curve in Fig. \ref{Fig:Experimental Results - video - PSNR-rate curves}.
The second strategy implemented the proposed method (Algorithm \ref{Algorithm:Proposed Method}) as the compression procedure that utilizes the HEVC standard. This approach was evaluated for a range of HEVC quality parameters to provide the corresponding PSNR-bitrate curve in Fig. \ref{Fig:Experimental Results - video - PSNR-rate curves}. The implementation of our iterative method (Algorithm \ref{Algorithm:Proposed Method}) run a maximum of 10 iterations or until convergence is detected.

\begin{figure*}[]
	\centering
	{\subfloat[{Source}]{\label{fig:stockholm_source_frame3}\includegraphics[width=0.32\textwidth]{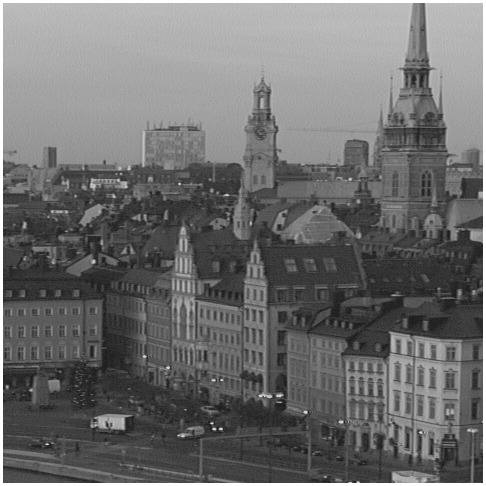}}}
	{\subfloat[{Regular}]{\label{fig:stockholm_regular_frame3__2_37bpp__28_29dB}\includegraphics[width=0.32\textwidth]{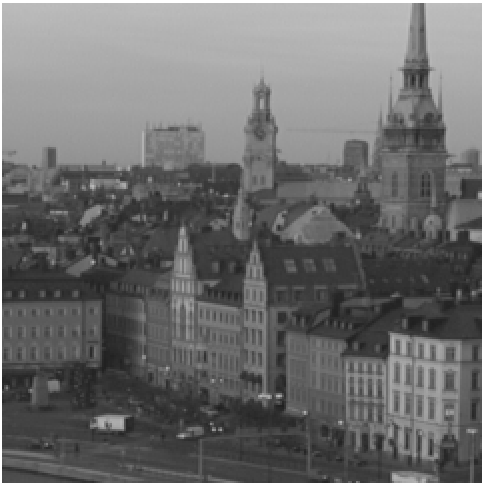}}}
	{\subfloat[{Proposed}]{\label{fig:stockholm_proposed_frame3__1_34bpp__29_45dB}\includegraphics[width=0.32\textwidth]{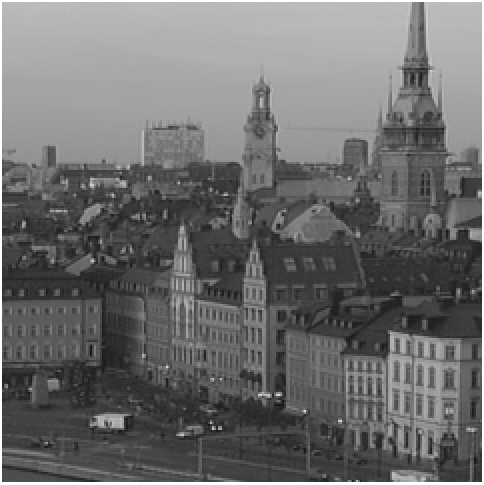}}}
	\caption{Coding a group of 10 frame from the 'Stockholm' sequence (spatial portion of 480x480 pixels). (a) The third source frame. (b) the rendered frame using regular compression (28.29 dB at 2.37 bpp). (c) the rendered frame using the proposed compression (29.45 dB at 1.34 bpp). } 
	\label{Fig:Experiments - Video - Visual Results - Stockholm}
\end{figure*}

\begin{figure*}[]
	\centering
	{\subfloat[{Source}]{\label{fig:shields_source_frame3}\includegraphics[width=0.32\textwidth]{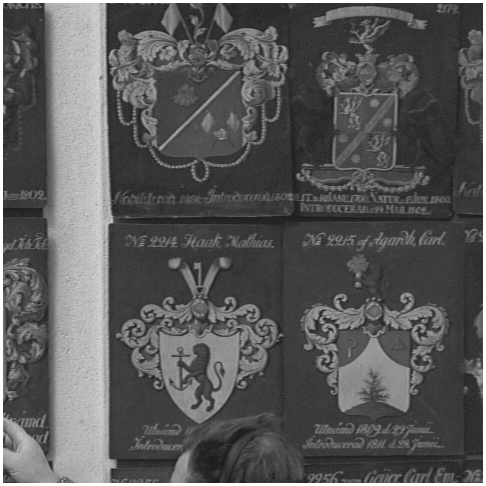}}}
	{\subfloat[{Regular}]{\label{fig:shields_regular_frame3__2_41bpp__27_93dB}\includegraphics[width=0.32\textwidth]{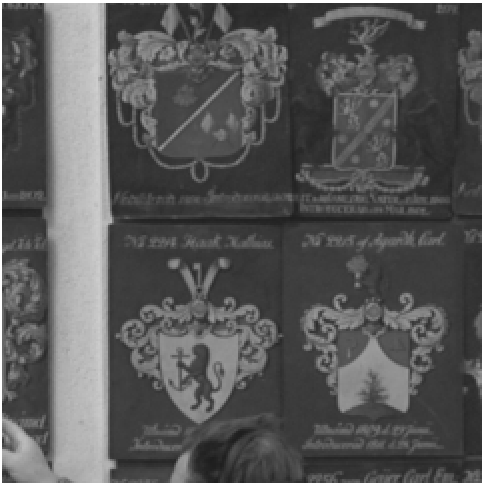}}}
	{\subfloat[{Proposed}]{\label{fig:shields_proposed_frame3__1_31bpp__29_31dB}\includegraphics[width=0.32\textwidth]{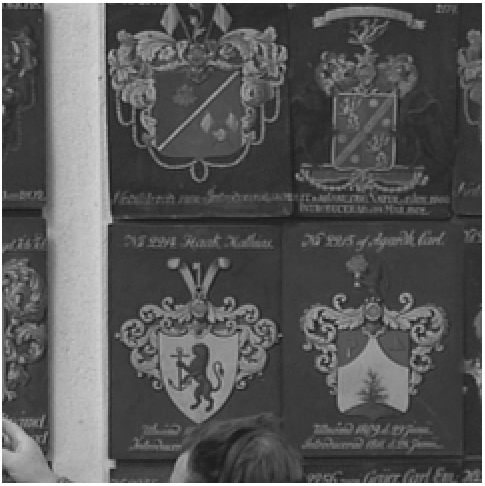}}}
	\caption{Coding a group of 10 frame from the 'Shields' sequence (spatial portion of 480x480 pixels). (a) The third source frame. (b) the rendered frame using regular compression (27.93 dB at 2.41 bpp). (c) the rendered frame using the proposed compression (29.31 dB at 1.31 bpp). } 
	\label{Fig:Experiments - Video - Visual Results - Shields}
\end{figure*}

In Fig. \ref{Fig:Experimental Results - video - PSNR-rate curves} we presented the PSNR-bitrate curves for the compression of segments of two video signals: 'Stockholm' and 'Shields'. Here, in Figures \ref{Fig:Experiments - Video - Visual Results - Stockholm} and \ref{Fig:Experiments - Video - Visual Results - Shields}, we show the third frame from the sequence, in its original form and in its rendered form using the regular approach and via the proposed method. Clearly, our method provides a more vivid image (also having higher PSNR) at a lower bit-rate.

\end{document}